\documentclass[12pt,preprint]{aastex}

\received{2005 September 22}
\accepted{2006 January 16}

\newcommand\iso[2]{$^{\rm #1}$#2}

\def\kmsec{\mbox{km~s$^{\rm -1}$}}
\def\teff{\mbox{T$_{\rm eff}$}}
\def\logg{\mbox{log $g$}}
\def\vt{\mbox{v$_{\rm t}$}}
\def\BmV0{\mbox{$(B-V)^{\rm 0}$}}
\def\VmK0{\mbox{$(V-K)^{\rm 0}$}}
\def\MV0{\mbox{$M_{\rm V}^{\rm 0}$}}

\def\etal{\mbox{et al.}}
\def\eg{\mbox{e.g.}}
\def\ie{\mbox{i.e.}}

\begin{document}

\title{
 Manganese Abundances in Cluster and Field Stars}

\author{
Jennifer S. Sobeck\altaffilmark{1},
Inese I. Ivans\altaffilmark{2, 3},
Jennifer A. Simmerer\altaffilmark{1},
Christopher Sneden\altaffilmark{1},\\ 
Peter Hoeflich\altaffilmark{1},
Jon P. Fulbright\altaffilmark{2},
and
Robert P. Kraft\altaffilmark{4}
}

\altaffiltext{1}{Department of Astronomy and McDonald Observatory,
University of Texas, Austin, TX 78712; jsobeck@astro.as.utexas.edu, jensim@astro.as.utexas.edu,
chris@verdi.as.utexas.edu, pah@astro.as.utexas.edu}

\altaffiltext{2}{The Observatories of the Carnegie Institution of
       Washington, 813 Santa Barbara St., Pasadena, CA 91101;
       iii@ociw.edu, jfulb@ociw.edu}

\altaffiltext{3}{Princeton University Observatory, Peyton Hall, Princeton,
       NJ 08544}

\altaffiltext{4}{UCO/Lick Observatory, Department of Astronomy and Astrophysics, 
University of California, Santa Cruz, CA 95064; kraft@ucolick.org}

\begin{abstract}

We have derived Mn abundances for more than 200 stars in 19 globular clusters.
In addition, Mn abundance determinations have been made for a comparable number of 
halo field and disk stars possessing an overlapping range of metallicities and stellar parameters. Our primary data
set was comprised of high resolution spectra previously acquired at the McDonald, Lick and Keck Observatories.
To enlarge our data pool, we acquired globular and open cluster spectra from several other 
investigators. Data were analyzed using synthetic spectra of the 6000 \AA\ Mn I triplet. 
Hyperfine structure parameters were included in the synthetic spectra computations.  Our analysis 
shows that for the metallicity range $-$0.7$>$[Fe/H]$>$$-$2.7 stars of 19 globular
clusters have a a mean relative abundance of $<$[Mn/Fe]$>$= $-0.37\pm0.01$ ($\sigma$ = 0.10), a value 
in agreement with that of the field stars: $<$[Mn/Fe]$>$= $-0.36\pm0.01$ ($\sigma$ = 0.08). 
Despite the 2 orders of magnitude span in metallicity, the $<$[Mn/Fe]$>$ ratio remains constant in both 
stellar populations.  Our Mn abundance data indicate that there is no appreciable variation in the 
relative nucleosynthetic contribution from massive stars that undergo core-collapse supernovae and thus,
no significant change of the associated initial mass function in the specified metallicity range.
\end{abstract}


\section{INTRODUCTION}

Trends in element abundances are utilized to uncover the formation patterns and evolutionary 
history of the Galaxy.  Comparison of chemical compositions between different stellar populations is
essential to this endeavor.  The interconnection between halo field and globular cluster stars is of extreme 
interest since their metallicity ranges overlap (\eg\ Laird \etal\ 1988,\nocite{Lai88} and references therein).  Recent
general overviews of abundance trends in halo populations have been done by, \eg, McWilliam (1997\nocite{McW97})
and Gratton \etal\ (2004\nocite{Gra04}).

With the exception of $\omega$ Cen (\eg\ Norris \& Da Costa 1995\nocite{Nor95}), stars of 
individual globular clusters display monometallicity, i.e. members of a globular cluster
possess approximately the same [Fe/H] value.  Elements of the proton-capture group (C, N, O, Na, Mg, and Al)
exhibit large star-to-star abundance variations in most globular clusters, and these discrepancies are
inordinately large as compared to those seen in halo field stars (Gratton et al. 2004\nocite{Gra04}, and references therein).
In contrast, members of the $\alpha$-element (like Si, Ca, and Ti) and neutron-capture element (like Y, Ba, La, and Eu)
groups display similar abundance patterns in most globular cluster and halo field stars.  Likewise, the relative
abundances of several Fe-peak elements (notably Sc, V, Cr, and Ni) appear to be almost identical
in the two stellar populations.  The vast majority of the Fe-group members have roughly solar
abundance ratios with two exceptions: copper and manganese.  The relative abundance of Cu is
known to be exceedingly subsolar in metal-poor field stars (at metallicities [Fe/H]$<$ $-$2, [Cu/Fe]
approaches $-$1; Sneden et al. 1991a\nocite{Snea91}, Mishenina et al. 2002\nocite{Mis02}).
An analogous deficiency of Cu in globular cluster stars has recently been reported by Simmerer et al. 
(2003)\nocite{Sim03}.  In the two stellar groups, the trend of [Cu/Fe] with [Fe/H] is identical within the limit
of observational uncertainty. 

Mn also has an established abundance deficiency in metal-poor stars.
Helfer et al. (1960)\nocite{Hel60} and Wallerstein (1962)\nocite{Wal62} were the first to report sub-solar
Mn and in 1978, Beynon \nocite{Bey78} verified these initial observations.  Later, Gratton (1989)\nocite{Gra89} 
improved Mn abundance determinations by employing hyperfine structure (HFS) data from Booth et al. (1984) to derive  
$<$[Mn/Fe]$>$= $-0.34\pm0.06$ for stars of metallicity [Fe/H]$<$ $-1$.  Three factors have hindered Mn abundance 
determinations: the lack of adequate hyperfine structure computations, the uncertainty of damping parameter values, 
and the absence of available transitions in the red portion of the visible spectrum (Gratton 1989\nocite{Gra89}; 
Prochaska \& McWilliam 2000\nocite{Proa00}).  Several surveys of metal-poor field stars have derived 
highly accurate [Mn/Fe] values (Gratton \& Sneden 1991\nocite{Gra91}; McWilliam et al. 1995\nocite{McW95}; 
Johnson 2002\nocite{Joh02}; Francois et al. 2003\nocite{Fra03}; Cohen et al. (2004b\nocite{Cohb04}).  
However, a systematic and comprehensive study of Mn abundances in globular cluster stars has not yet taken place.      

In this paper we present Mn abundances for several hundred cluster and field stars in the 
metallicity range of 0.0$\gtrsim$[Fe/H]$\gtrsim$--2.7.  Our intent is two-fold: first, we want to 
ascertain whether globular cluster stars have the same Mn abundance as stars of the halo field; and second, we 
wish to confirm the Mn abundance trend across the entire metallicity spectrum, as well as across the stellar populations, in order
to further resolve the nucleosynthetic origin of this element.  In \S 2 we relay particulars about each data set
and characterize the general nature of the data.  A justification of line selection and a description of the analysis is found in \S 3.  
An account of all abundance values is given in \S 4.  Finally in \S 5, a discussion of these Mn results ensues.      

\section{OBSERVATIONS AND DATA REDUCTION}

In this study, Mn abundance measurements were made in three stellar populations: globular clusters, open clusters, 
and the halo field.  
Spectroscopic and equivalent width data were acquired from numerous sources.  A significant portion of 
the globular cluster and halo field data were gathered by the Lick-Texas group 
(LTG).  These LTG data constitute a basis set for our Mn abundance survey.
Table 1 lists the relevant observational parameters and literature sources for the LTG data.  Cluster sample 
size varies from as few as 2 to as many as 23 stars.  The two field star surveys each have a sample size in 
excess of 80 stars. Stars in the field data sample exist in a variety of evolutionary states whereas the bulk of the 
globular cluster data are red giants.  Three facilities were used for the LTG observations: the Keck I 10.0 m telescope 
equipped with the High Resolution Echelle Spectrometer (HIRES; Vogt \etal\ 1994\nocite{Vog94}), 
the Lick 3.0 m telescope equipped with the Hamilton spectrograph (Vogt 1987\nocite{Vog87}), and 
the McDonald 2.7 m telescope equipped with the ``2d-coud\'{e}'' spectrograph (Tull \etal\ 1995\nocite{Tul95}).  
For the various instrument configurations, the resolution (R $\equiv$ $\lambda$/$\Delta\lambda$) ranges from 30,000 to 60,000, and 
the estimated signal-to-noise ratio (S/N) varies between 25 and 180. 
The software packages IRAF\footnote{IRAF is distributed by the 
National Optical Astronomy Observatories, which are operated by the Association of
Universities for Research in Astronomy, Inc., under cooperative agreement
with the National Science Foundation.} and SPECTRE (Fitzpatrick \& Sneden 1987\nocite{Fit87}) were used for standard data reduction 
processes such as bias and flat-field correction, order extraction, cosmic ray elimination, continuum adjustment, 
and wavelength correction.  

The remainder of the globular and open cluster spectra were obtained from several external sources.  Data contributors,
as well as observational details, are found in Table 2.  These data were collected at several facilities: the 
Very Large Telescope (VLT), Apache Point Observatory (APO), Cerro Tololo Inter-American Observatory (CTIO), and Keck. 
The various telescope-spectrograph combinations yielded resolutions of 24,000$\leq$R$\leq$60,000 and S/N values
between 30 and 135.  A variety of data reduction and analysis programs were used by the contributors, and for further
details the reader should consult the original references (as listed in Table 2). 

\section{ANALYSIS}

Line selection was based on metallicity and effective temperature parameters.  A considerable number of the 
globular cluster and halo field stars in the data sample have \teff$\leq$ 4850.  Accordingly, analysis of 
the strong Mn lines at 4030 \AA\ and 4823 \AA\ was not feasible due to the flux constraints of the data and the 
probable saturation of these features.  In addition, most extant cluster spectra do not extend to the blue-violet wavelength region. 

\begin{figure}
\epsscale{0.80}
\plotone{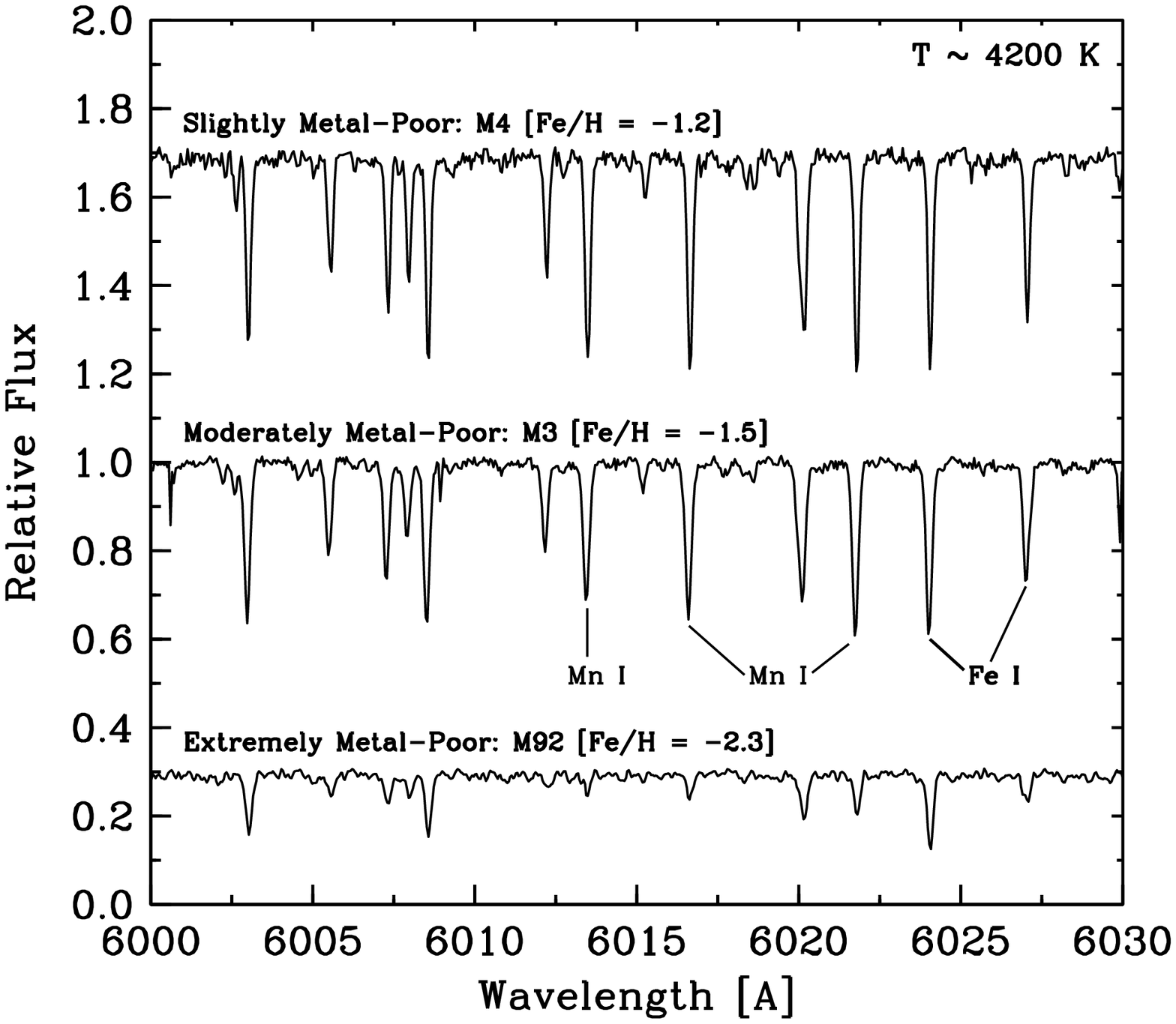}
\caption{Comparison of the spectra from globular clusters of differing metallicities.  Note that temperature is roughly the same 
         for all of the spectra.  The three Mn and two Fe lines used in the abundance analysis are indicated in 
         this figure.  As metallicity decreases, some of the spectral features become undetectable.
\label{f1}}
\end{figure}

To ascertain Mn abundance in these stars, a wavelength range of 6000-6030 \AA\ was chosen, which encompasses three 
Mn I spectral features at 6013.51, 6016.64, and 6021.82 \AA.  
These lines are the sole strong transitions of Mn in the yellow-red 
spectral regime.  Two Fe lines at 6024.06 and 6027.05 \AA\, which are roughly of the same excitation potential as the Mn features,
were employed for a local iron abundance determination.  The use of these nearby Fe transitions eliminates 
possible discrepancies in continuous opacity and issues with regard to first-order continuum placement.
And although the convenience of these two lines must not be understated (as they lie on the same spectral
order as the three Mn features), our goal was to obtain local [Fe/H] values for the [Mn/Fe] determination, not to 
replace the more extensive [Fe/H] assessments done in previous LTG studies.  Figure~\ref{f1} features all 
of the lines used for analysis and roughly illustrates line strength as a function of metallicity.
 
\subsection{Model Atmospheres and Techniques}

For the LTG data, we adopted the stellar atmospheric parameters as reported by the original papers. 
We employed the model stellar atmospheres that were generated for those papers from the 
MARCS (Gustaffson et al. 1975\nocite{Gus75}) and  ATLAS (Kurucz 1993\nocite{Kur93}) codes.  Table 3 presents the \teff, 
\vt, and \logg~numbers for the LTG data set.  Model atmospheres did not normally accompany the data from outside sources.  
We took the stellar atmospheric parameters as reported by the contributors and generated the models for
these quantities from the grid of ATLAS models without convective overshoot (Castelli et al. 1997\nocite{Cas97}) using
software originally provided by A. McWilliam.  Table 4 displays the parameters for the stars of the external source data set.  

In order to refine the line list, we synthesized a portion of the solar spectrum (6000-6030 \AA).
The observed center-of-disk photospheric spectrum is that of Delbouille et al. (1990\nocite{Deb90})
\footnote{We employed the electronic version available on the website of the Base de donnes 
Solaire Sol, \url{http://bass2000.obspm.fr/home.php}}.  
We selected a Holweger-M\"{u}ller model with a microturbulent velocity of \vt= 0.80\kmsec, 
a value in accord with other solar abundance surveys 
(Holweger \& M\"{u}ller 1974\nocite{Hol74}; Grevesse \& Sauval 1999\nocite{Gre99}).  
We used the standard LTG value of log$\epsilon$(Fe)$_{\sun}$= 7.52 as set by Sneden \etal\ (1991b)\nocite{Sneb91}.  
The initial basis for this value originates from the work of Anders \& Grevesse (1989)\nocite{And89}.  
Further confirmation of this value was done by Anstee et al. (1997\nocite{Ans97}), who used neutral iron lines 
to derive an iron abundance for the Sun of 7.51$\pm$0.01. Several other studies arrive at approximately the same value 
(to within 0.1 dex) for the solar photospheric abundance of iron (e.g. Raassen \& Uylings 1998\nocite{Raa98}; Asplund et al.
2000\nocite{Asp00}).  We also adopt log$\epsilon$(Mn)$_{\sun}$= 5.39 as recommended by Anders \& Grevesse (1989\nocite{And89}).
Note, however, that there is a significant discrepancy between the solar photospheric and meteoritic CI chondrite 
[log$\epsilon$(Mn)$_{meteor}$= 5.50; Lodders 2003\nocite{Lod03}] values for Mn.    

\begin{figure}
\epsscale{0.80}
\plotone{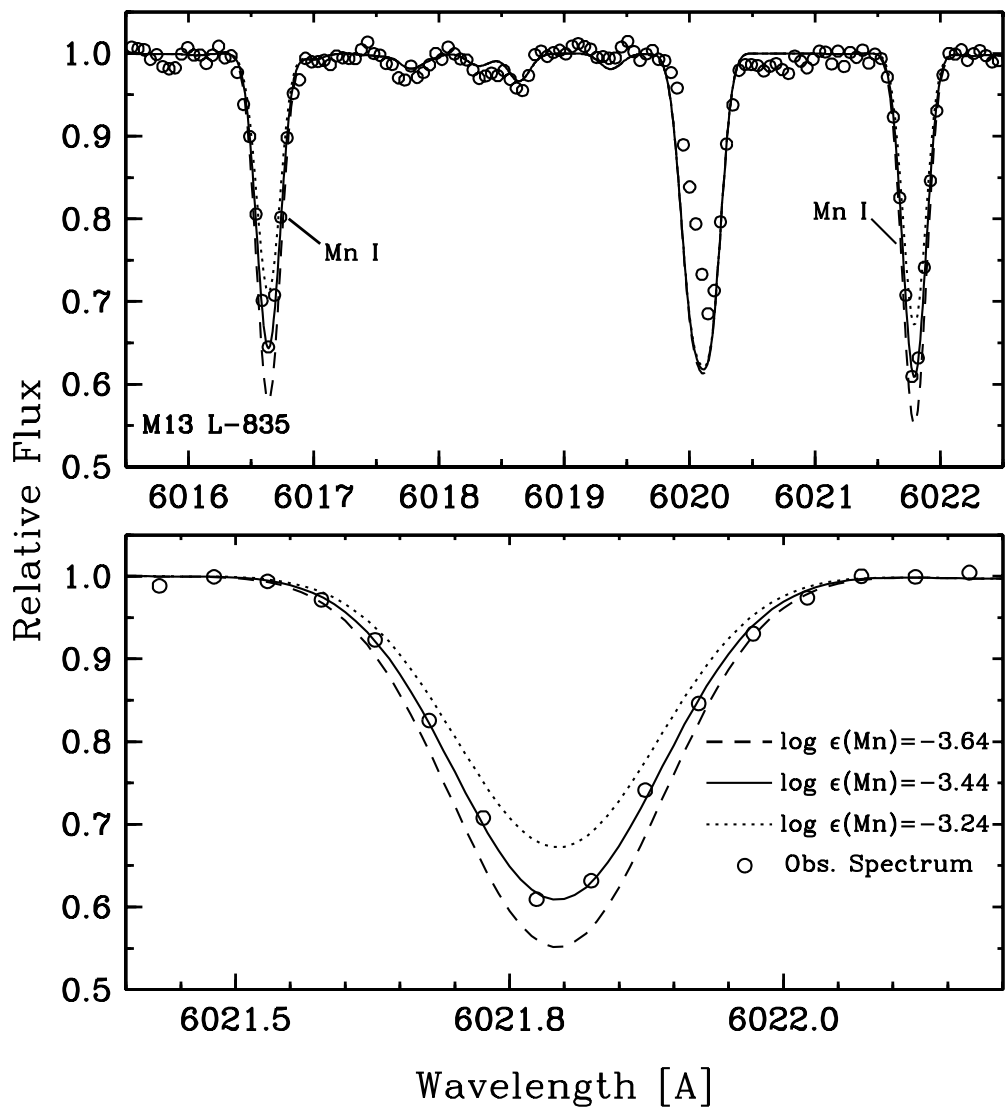}
\caption{Comparison of the synthetic and observed spectrum for one M13 star. The top panel
         displays the observed and synthetic spectra for a wavelength range that encompasses the 
	 6016 and 6021 \AA\ Mn lines.  The bottom panel focuses on the 6021 \AA\ Mn feature and 
	 highlights the effects of incremental changes in abundance.  Changes as small as 0.2 dex cause 
	 distinct variation in the synthesized spectrum.
\label{f2}}
\end{figure}

Spectrum synthesis was employed to determine the abundances as accurate determinations from transitions with
multiple HFS components necessitate this technique.  Abundance derivations that rely
solely on the measurement of equivalent width values do not properly account for lines containing
HFS without the introduction of an artifact (i.e., an arbitrary increase in microturbulent velocity; Cohen 1978\nocite{Coh78}).  
To generate synthetic spectra and to calculate abundances, the current version of the LTE line analysis code MOOG 
(Sneden 1973)\nocite{Sne73} was used.  The raw relative flux values generated by this code were convolved with Gaussian 
broadening functions to reproduce the combined effects of astrophysical (\ie\, macroturbulence) and 
instrumental (\ie\, spectrograph slit) origin.   Figure~\ref{f2} shows a sample spectrum synthesis.  
In cases in which the spectra were not available, we employed literature values of the equivalent width
measurements.  In those instances we computed synthetic spectrum fluxes, which were then summed
to force-fit the observed equivalent width values.  This technique was verified in some spectra for
which synthetic fits were also made to observed spectra.

\subsection{Line Parameters}

Two Fe I features (6024 and 6027 \AA) are available for abundance determinations in the specified wavelength range.
A reliable Fe abundance may be obtained from these neutral lines as their excitation potential is large 
($\chi$$>$4.0 eV); consequently, they are not as susceptible to temperature effects 
and departures from LTE (Grevesse et al. 1996\nocite{Gre96}).  
Multiple literature sources give a transition probability 
for the 6027 \AA\ feature.  The emission measurement technique of O'Brian et al. (1991) \nocite{Obr91} yielded a gf-value 
for the 6027 \AA\ line that is in good agreement with that found by the absorption line technique of 
Blackwell et al. (1982) \nocite{Bla82}.  We adopted the O'Brian et al. log (gf) value for this line.  

Unfortunately, neither O'Brian et al. (1991) nor Blackwell et al. (1982) give a transition
probability for the 6024 \AA\ feature. Literature sources for this line include the early work of 
Wolnik et al. (1970), log (gf)= $-$0.06$\pm$0.00
\nocite{Wol70}; the solar line inversion value of Th\'{e}venin (1990), log (gf)= $-$0.02$\pm$0.02\nocite{The90}; and 
the semi-empirical 
derivation of Kurucz (1993), log (gf)= $-$0.120\nocite{Kur93}.  Taking into consideration the lack of modern 
laboratory atomic physics input into these numbers, we opted to perform an empirical derivation of the 6024 \AA\ gf-value.  
An initial line list (in the specified 30 \AA\ wavelength range) was assembled from Kurucz (1993)\nocite{Kur93} data.  
A synthetic spectrum was generated from this list and compared 
to the observed solar spectrum.  Modification of the line list (\ie\, revision of gf-values and deletion of non-essential
features) occurred until the difference between the observed spectrum and the synthetic spectrum was minimized.  
With the refined line list in place, the iterative determination of 
the 6024 \AA\ gf-value proceeded.  The abundances of Mn and Fe were set to their corresponding solar values and the   
smoothing and continuum were fixed. Then the transition probability and the van der Waals damping parameter (C6) 
of the 6024 \AA\ line were allowed to vary until a good fit was achieved.  
A final value of log(gf)$_{6024}$ = 0.04 was obtained, with associated 
enhancement of the C6 damping parameter of E$\gamma$= 2.2.  The result for the damping parameter enhancement
is in agreement with the finding of Anstee et al. (1997\nocite{Ans97} and references therein) that lines with an excitation
potential greater than 3.0 eV generally have an E$_\gamma$$> 2.1$. 
            
 With a nuclear spin of $I = 5/2$ and a magnetic dipole moment of $\mu$$_{I}$$= 3.4687$~$\mu$$_{N}$ 
(Lederer \& Shirley 1978\nocite{Led78}), Mn has a sizable HFS.  The effect of HFS is to 
desaturate and broaden the lines of Mn.  The strongest transitions of Mn are particularly susceptible.  
To ensure the accurate computation of Mn abundance, HFS was taken into account. Oscillator strengths for the 6013 and 
6021 \AA~ Mn lines were taken from Booth et al. (1983, 1984\nocite{Bot83}\nocite{Bot84}).  Additional data were acquired
from the Kurucz (1993)\nocite{Kur93} line list.  Neither Booth et al. (1983\nocite{Bot83}) nor the NIST database 
\footnote{The associated NIST website is: 
\url{http://physics.nist.gov/PhysRefData/ASD/index.html}.} (Martin et al. 1999\nocite{Mar99}) give 
a transition probability for the 6016 \AA~line.  As before, the gf value for the 6016 \AA\ Mn 
line was determined iteratively via a fit to the observed solar spectrum.  Notably, the 6016 line
possesses a significant Fe contaminant, whereas the 6013 and 6021 \AA\ features do not contain any prominent blends.  
So, little weight is accorded to the abundance derived from the 6016 \AA\ feature due to line contamination
and slight uncertainty in oscillator strength value (it is used for a consistency check only).  
Final transition probabilities for all lines are reported in Table 5. 

\section{RESULTS}
\begin{figure}
\epsscale{0.80}
\plotone{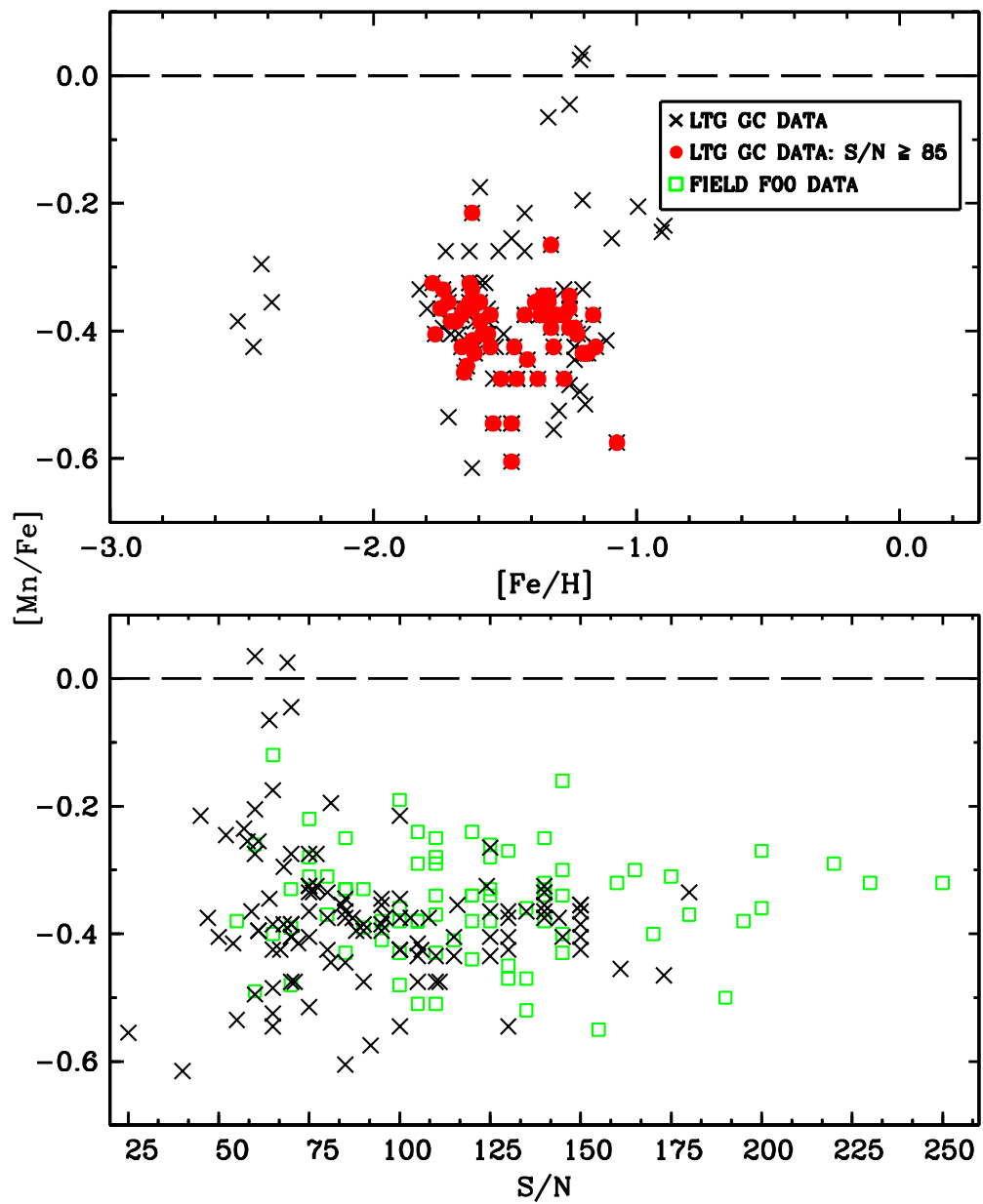}
\caption{Interdependence of [Mn/Fe] and S/N for the LTG data sample.  The top panel presents the entire S/N range of the LTG 
         globular cluster data set as well as those points with S/N$>$85.  The bottom panel illustrates the correlation
	 of Mn abundances with S/N for LTG globular cluster and field data in the metallicity range --0.7$>$[Fe/H]$>$--2.7. 
\label{f4}}
\end{figure}

\begin{figure}
\epsscale{0.80}
\plotone{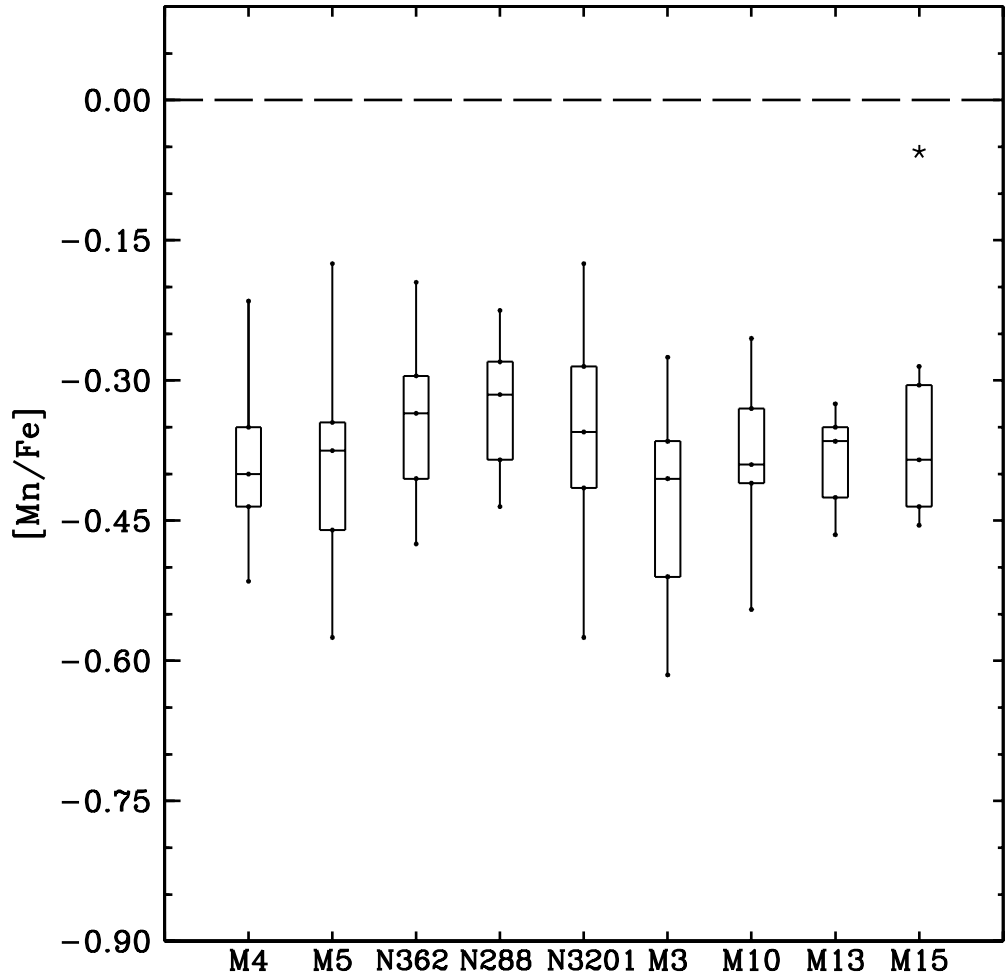}
\caption{Box plots for nine representative globular clusters. 
         For each cluster, the boxed region encompasses the interquartile (middle 50\%) of its [Mn/Fe] data.  Also featured are 
         the median (horizontal line), range (vertical lines; excludes outliers), and outliers (an outlier has
	 a value greater than 1.5 times the interquartile range).  The ordering of the clusters is in decreasing [Fe/H]$_{avg}$.
\label{f5}}
\end{figure}

\begin{figure}
\epsscale{0.80}
\plotone{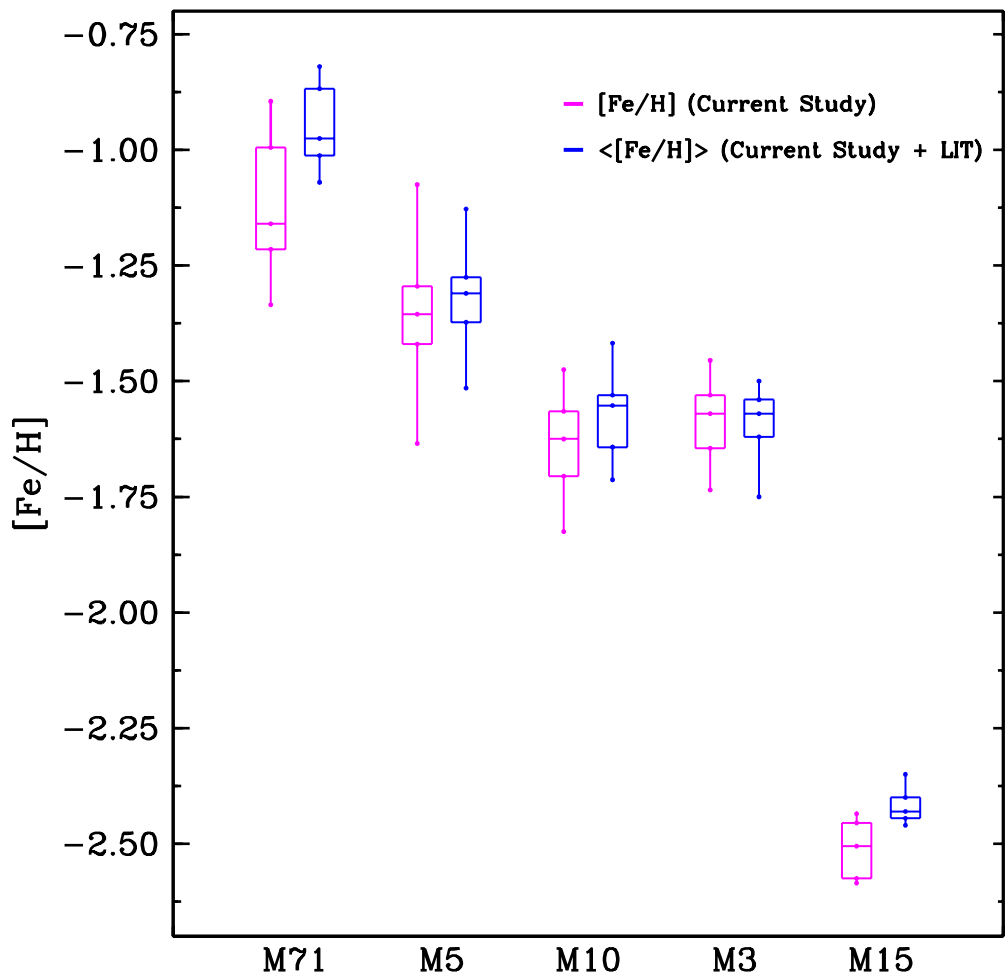}
\caption{Box plots for selected globular clusters.  For a few cases, averaging our derived [Fe/H] values with those 
         reported in the literature serves to reduce the spread in metallicity.  As a consequence of this averaging process, 
	 the median [Fe/H] value for the cluster increases. In general, marginal benefit is gained from the averaging 
	 process (as clearly illustrated by M3).  
\label{f6}}
\end{figure}

Our essential finding is that in the metallicity range $-$0.7$>$[Fe/H]$>$$-$2.7 the Mn abundances in globular cluster
stars are equivalent to those of halo field stars. Figure~\ref{f3} displays the [Mn/Fe] ratio as a function of [Fe/H] for all data.  
The mean abundance in the specified metallicity range is $<$[Mn/Fe]$>$= $-$0.37$\pm$0.01 ($\sigma$ = 0.10) for  
globular cluster stars and $<$[Mn/Fe]$>$= $-$0.36$\pm$0.01 ($\sigma$ = 0.08) for halo field stars. Figure~\ref{f4} 
presents the correlation of S/N with [Mn/Fe] for the LTG data set.  As shown in the
bottom panel, very high S/N data (S/N$> 175$) give an extremely consistent [Mn/Fe] value. In Figure~\ref{f5} 
the scatter in [Mn/Fe] is shown for selected globular clusters with large data samples.  Intra-cluster variations with respect to
Mn abundance are nominal, and scatter is within observational error.
However, our chosen Fe features contribute to scatter in Fe, as demonstrated in Figure~\ref{f6}.  In a few cases 
the spread in metallicity is larger than 0.3 dex.  The inclusion of more Fe lines (of both ionization states) 
would somewhat improve the abundance determination.  So for the LTG clusters, we list our 
[Fe/H] values, as well as our [Fe/H] ratios averaged with those reported by the original reference.
Table 6 presents the [Fe/H] and [Mn/Fe] values that result from this averaging process.  In the designated [Fe/H]
range, we were able to obtain from the literature [Mn/Fe] data points for five globular clusters: M55, M68, 
NGC 104, M71, and M30.  These literature [Mn/Fe] values are in fairly 
good agreement with our own.  A few clusters in our sample were a bit problematic, and we discuss these 
clusters in the following sections.  We also address the noticeable data gaps in the extremely-poor
metallicity range ([Fe/H]$<$ $-$2.7 dex) and the slightly-metal poor range ([Fe/H]$>$ $-$0.7 dex) in section \S4.4.

\subsection{Error Analysis}

Four main factors contribute to possible errors in our abundances: choice of model, sensitivity to stellar parameters, 
quality of observational data, and modification of elements of the spectral fit process.  
To assess the ramifications of model/parameter variation across the entire data set, we studied representative stars of three 
metallicity classes: slightly metal-poor (SMP), moderately metal-poor (MMP), and extremely metal-poor (EMP). 
The selection of the stellar atmosphere model (be it MARCS or Kurucz) seemed to have little effect on either [Mn/H] or [Fe/H] 
(with a maximum change of 0.07 dex in [Fe/H] for a SMP star).
The relative abundances are not very responsive to slight changes in the stellar
parameters.  For a change of $\pm$100 K in \teff, the largest effect was seen in the [Fe/H] ($\pm$0.10 dex) of SMP stars.
An alteration in the \logg\ value of $\pm$0.20 dex had a maximum response in the [Fe/H] of EMP stars with a change of
$\pm$0.10 dex.  The [Mn/H] value responded similarly, but taken together in the ratio [Mn/Fe]
the effect cancels out.  And for $\Delta$\vt= $\pm$0.20\kmsec, the greatest change is seen in EMP stars
with $\pm$0.09 dex in both [Mn/H] and [Fe/H].  Overall, the abundance error from the variation of these stellar parameters
does not exceed $\pm$0.10 dex.       

The S/N across the entire data set did vary by a substantial amount: 25$\le$S/N$\leq$180.
 For data of generally high quality (S/N$> 75$), the abundance determined via spectral synthesis fit is 
good to within $\pm$0.05 dex.  Conversely, the fit for low quality data is not as solid and may fluctuate 
by as much as $\pm$0.10 dex.  Further considerations are continuum normalization and smoothing parameters of the fit.  
Placement of the continuum might affect the fit by as much as $\pm$0.03 dex, whereas alteration of the FWHM of the 
fitting function (normally a Gaussian for most stars) may result in an abundance change of roughly $\pm$0.05 dex.     

Non-LTE effects should also be taken into consideration.  For metal-poor stars, overionization (and its impact
on surface gravity) is indeed a factor (Thevenin \& Idiart 1999\nocite{The99}), but to what degree is
not clear (Kraft \& Ivans 2003\nocite{Kra03}; Korn 2004\nocite{Kor04}).  To date, no non-LTE Mn abundance 
calculations have been published for stars of any type.  In a survey of 
metal-poor giants, Johnson (2002)\nocite{Joh02} attempted to quantify the effect on Mn by estimating a non-LTE 
\logg.  Johnson demonstrated that modification of the \logg~value elicited a change of roughly --0.10 dex in Mn abundance.  
Ivans et al. (2001)\nocite{Iva01} suggest that as long as the abundance ratio consists of 
two neutral species (as is the case in our study) the relative non-LTE effects are minimized.

\subsection{M71}

Our initial result for M71 indicated a high Mn abundance as compared to other
globular clusters in our data sample.  For 10 stars, we derived $<$[Mn/Fe]$>$= $-$0.16 ($\sigma$ = 0.14) with an
average  metallicity of $<$[Fe/H]$>$= $-$1.12 ($\sigma$ = 0.15).  The data have an unusually large scatter in 
both Mn and Fe.  We must take into consideration the fact that our M71 observational runs at the Lick 3.0 m
telescope occurred in 1989 and 1991, prior to the update of the echelle spectrograph.
If we discount the four most anomalous data points (which correspond to the lowest S/N values), 
then  the $<$[Fe/H]$>$ for M71 becomes $-$1.04 ($\sigma$ = 0.12) and the $<$[Mn/Fe]$>$ is   
$-$0.26 ($\sigma$ = 0.08).  Also, if we average our [Fe/H] ratios with those reported in the original LTG M71 study, then
$<$[Fe/H]$>$= $-$0.91 ($\sigma$ = 0.06) and $<$[Mn/Fe]$>$= $-$0.38 ($\sigma$ = 0.11).
With regard to these considerations, the M71 abundances are much more in line with other data points of similar
metallicity.  Using Keck I data acquired in 2002, Ramirez \& Cohen \nocite{Ram02} were able to ascertain Mn abundances for M71.  
For this cluster, they derive $<$[Fe I/H]$>$= $-$0.71$\pm$0.08 and $<$[Mn/Fe]$>$= $-$0.27$\pm$0.11.  
Due to the higher resolution and S/N of the Ramirez \& Cohen (2002\nocite{Ram02}) data, their abundance
values are to be preferred (Figure~\ref{f3}, bottom).

\subsection{Comparison of Cluster Results: NGC 6528 and C261}

We are able to compare our derived [Mn/Fe] ratios to literature values for two clusters of high
metallicity, NGC 6528 and Collinder 261 (Cr 261).  NGC 6528
presents an opportunity to study the cluster populations of the Galactic
Bulge.  It lies in Baade's window and thus has only
moderate reddening.  Although Cr 261 is an open cluster, it may be likened to globular clusters,
as it is similar in age (roughly 9 Gyr; Janes \& Phelps (1994)\nocite{Jan94}).

For three red horizontal branch stars of NGC 6528, Carretta et al. (2001)\nocite{Car01} 
found \\$<$[Fe I/H]$>$= 0.07 ($\sigma$ = 0.02) and $<$[Mn/Fe]$>$= $-$0.37 ($\sigma$ = 0.07).  
In our examination of three different stars from this cluster,
we derive mean values of $<$[Fe/H]$>$= $-$0.24 ($\sigma$ = 0.19) and $<$[Mn/Fe]$>$= $-$0.25 ($\sigma$ = 0.06).  As Zoccali et al. 
(2004\nocite{Zoc04}) have pointed out in their study of NGC 6528, factors that affect abundance derivations include 
effective temperature assessment (both spectroscopically and photometrically derived 
parameters contain inherent errors) and continuum determination (placement of the continuum may be largely 
variable due to the presence of molecular bands and $\alpha$ enhancement).  
Our Fe values for this cluster do show a large spread: -0.37$\leq$[Fe/H]$\leq$-0.03
(the temperature range of the sample stars is a likely factor).  Also, special attention should be paid to
the broadening factors used in abundance determination (Zoccali et al. 2004\nocite{Zoc04}).  While taking 
into consideration all the issues mentioned above, we remark that we still find a substantial underabundance
of Mn in NGC 6528.

Carretta et al. (2005\nocite{Car05}) also observed six red clump and red giant branch stars in Cr 261.  
For this cluster, they found $<$[Fe I/H]$>$= $-0.03$ ($\sigma$ = 0.04) and $<$[Mn/Fe]$>$= $-0.03$ ($\sigma$ = 0.04).
We employed a different data set (Friel et al. 2003\nocite{Fri03}) that contains four 
of the stars that were in the Carretta et al. (2005) sample.  
Our analysis of Cr 261 giants yields $<$[Fe/H]$>$= $-$0.36 ($\sigma$ = 0.21) and $<$[Mn/Fe]$>$= $-$0.32 ($\sigma$ = 0.13).  Data concerns
might include instrument resolution and S/N values.  Moreover, there is definite
sensitivity in the data to the selection of \vt, transition probabilities, and \logg\ values (Carretta et al. 2005).
We note that there is significant scatter in our Fe abundance, and it is indeed a rather low value.  
In both studies, one target star gave consistently low [Fe/H] and [Mn/Fe] values as compared to other stars in the data set.
None of the studies chose to exclude this star (most likely due to the small data sample for Cr 261). These are preliminary 
investigations of clusters in the metallicity regime [Fe/H]$> -$0.70 and the acquisition of more data in this range 
will be necessary.  Future efforts will also focus on open cluster abundances.

\subsection{Other Mn Abundance Analyses}

Several investigations of the Mn abundance ratio have been done in various metallicity regimes
and stellar populations.  We briefly detail some of those here along with the associated [Mn/Fe] results.
For field stars of low metallicity ([Fe/H]$<$ $-$1.7), Johnson (2002\nocite{Joh02}) obtained a 
subsolar Mn abundance. Studies by Cohen et al. (2004\nocite{Cohb04}) and Francois et al. (2003\nocite{Fra03})
find that Mn decreases steadily below metallicity [Fe/H]$\sim$ $-$3.0.

Examinations of Mn in metal-rich field stars are plentiful in the literature.  Solar neighborhood stars in the range 
-0.15$<$[Fe/H]$<$0.45 have been found by Chen et al. (2003\nocite{Che03}) to possess a relatively constant [Mn/Fe] ratio 
hovering roughly at zero. Alternatively, the solar neighborhood survey of Allende Prieto et al. (2004\nocite{All04}) 
reported that [Mn/Fe] rises in step with [Fe/H]. Mn abundance determinations of the disk field stars 
include Feltzing \& Gustafsson (1998\nocite{Fel98}), Prochaska et al. (2000b\nocite{Prob00}), and 
Reddy et al. (2003)\nocite{Red03}. Generally, these studies find that as [Fe/H] approaches zero,
so, accordingly, does the Mn abundance with respect to Fe 
(with the rough determination of the solar Mn abundance level at solar metallicity).  
In addition, these studies report that above [Fe/H]$=$ 0, increases in [Fe/H] correspond to attendant
increases in [Mn/Fe]. E. Carretta et al. (2006, in preparation\nocite{Car06}) have 
conducted Mn abundance analyses of several open clusters. 
They have found that the open clusters of their data sample do mimic the trend of the disk.  
Now, Prochaska et al (2000\nocite{Prob00}) contend that Mn abundance differs between the thick disk 
and the thin disk.  They conclude that Mn in the thick disk is normally underabundant with respect to
the thin disk.  This finding is being subjected to further scrutiny (Reddy et al. 2006 \nocite{Red06}).  

Bulge globular clusters have not been well analyzed and remain somewhat of a 
mystery (the notable exception, of course, being NGC 6528).  
McWilliam et al. (2003\nocite{McW03}) have discovered that the [Mn/Fe] values 
of bulge giants follow the trend of disk stars.  In a separate study, McWilliam et al. (2003\nocite{McW03}) 
examined the [Mn/Fe] ratio in the Sagittarius dwarf spheroidal 
galaxy and found a fairly consistent underabundance with respect to the stars of the bulge and disk populations.  
In order to have a more complete view of these metal-rich stellar populations, further study is requisite.

\section{DISCUSSION AND CONCLUSIONS}

We have derived Mn abundances for hundreds of globular cluster, open cluster, and halo 
field stars.  We used spectral synthesis in order to obtain a [Fe/H] and [Mn/Fe]
ratio for each star.  In the range $-$0.7$>$[Fe/H]$>$$-$2.7, globular cluster stars exhibit 
a mean relative abundance of $<$[Mn/Fe]$>$= $-$0.37$\pm$0.01 ($\sigma$ = 0.10), which is 
the same (to within the levels of uncertainty) as that of halo field stars, $<$[Mn/Fe]$>$= $-$0.36$\pm$0.01 ($\sigma$ = 0.08).
There is no statistically significant difference with regard to Mn abundance between the halo field 
and globular clusters. 

\begin{figure}
\rotate
\epsscale{0.80}
\plotone{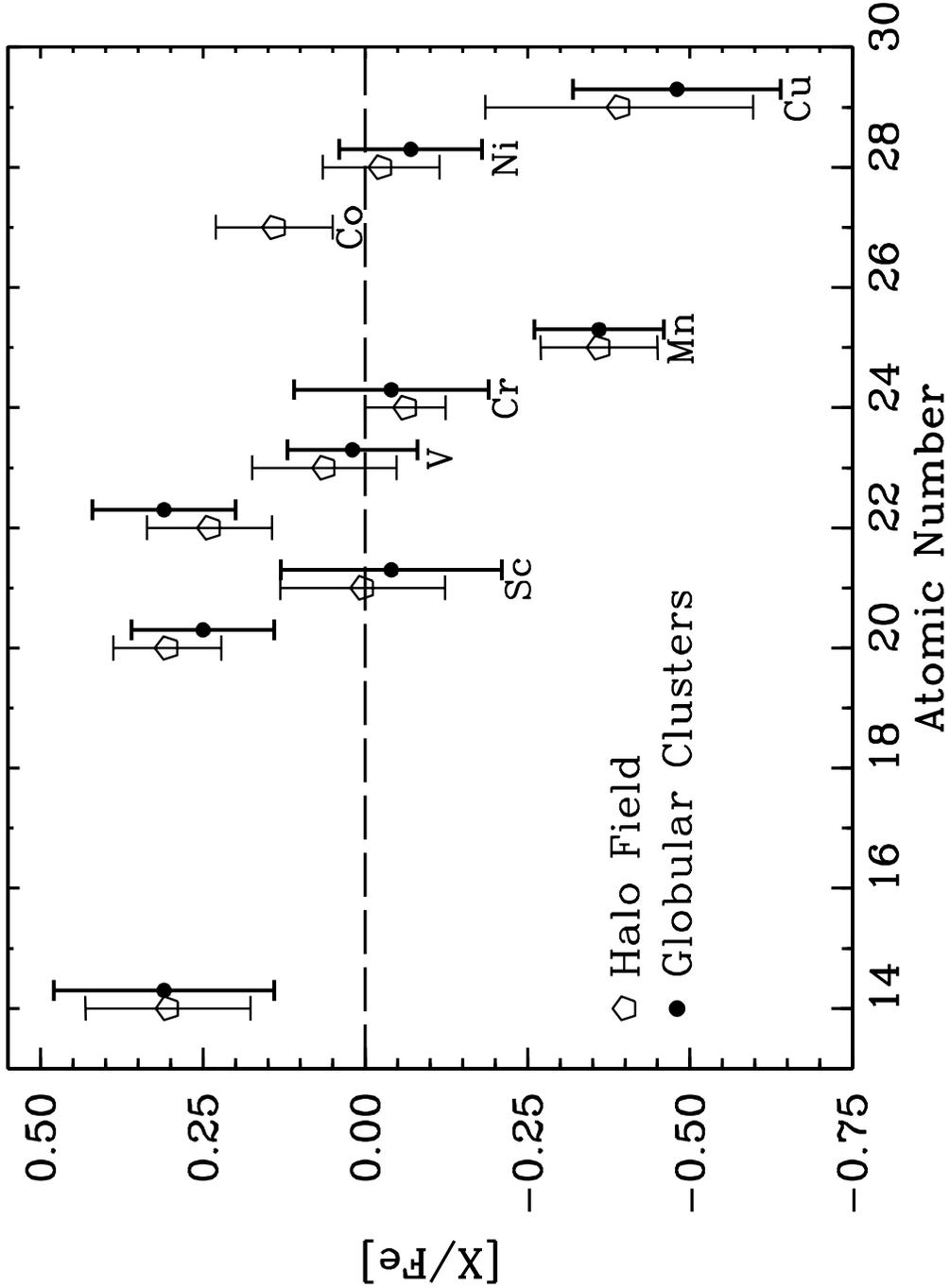}
\caption{Average abundance ratios with associated standard deviation values 
         for some of the Fe-peak and $\alpha$-elements in the range --0.7$>$[Fe/H]$>$--2.7.  Globular cluster data are 
	 obtained from the LTG references.  The Mn values are provided by the current study. 
         Data for halo field stars for every element 
	 except Cu and Co are taken from Gratton et al. (2003) and Fulbright (2000).  Cu field star data are obtained 
	 from Mishenina et al. (2002) and Co field star data are taken from Johnson (2002).  Interestingly, there is a lack of 
	 Co data for globular clusters.  Note that in almost every case, the average values for globular cluster and halo field 
         stars are roughly equal to one another.
\label{f7}}
\end{figure}

Figure~\ref{f7} displays the average abundance ratios of Fe-peak elements in halo field and globular cluster stars in the 
metallicity range $-$0.7$>$[Fe/H]$>$$-$2.7. Several points may be
gleaned from this plot. First, and most important, the elemental abundance ratios are equivalent in the 
two stellar populations.  Second, the relative abundances for many members of the Fe group (Sc, V, Cr, Co, and Ni) 
are roughly solar over this metallicity range.  And third, the abundances of a few odd Z-numbered elements 
(namely, Mn and Cu) are deficient with respect to their even Z-numbered Fe-peak counterparts.  

Nucleosynthesis of Mn occurs primarily via decay of \iso{55}{Co} (Nakamura et al. 1999\nocite{Nak99}).  
Another possible nucleosynthetic pathway for Mn is $\alpha$-capture by \iso{51}{V}.  
The main site for Mn formation is the incomplete explosive Si-burning region (Nakamura et al. 1999\nocite{Nak99}).
In the metallicity range of interest ($-$0.7$>$[Fe/H]$>$$-$2.7), core-collapse supernovae (SNe)
are predominantly responsible for the production of Mn.  Yields of Mn rely heavily upon the 
neutron excess (Umeda \& Nomoto 2002\nocite{Ume02}).  The [Mn/Fe] ratio
depends on the mass cut (as Fe has two production sites: the incomplete and complete Si-burning regions) and the
explosion energy (with little dependence on stellar mass; Umeda \& Nomoto 2002\nocite{Ume02}). 
  
\begin{figure}
\epsscale{0.80}
\plotone{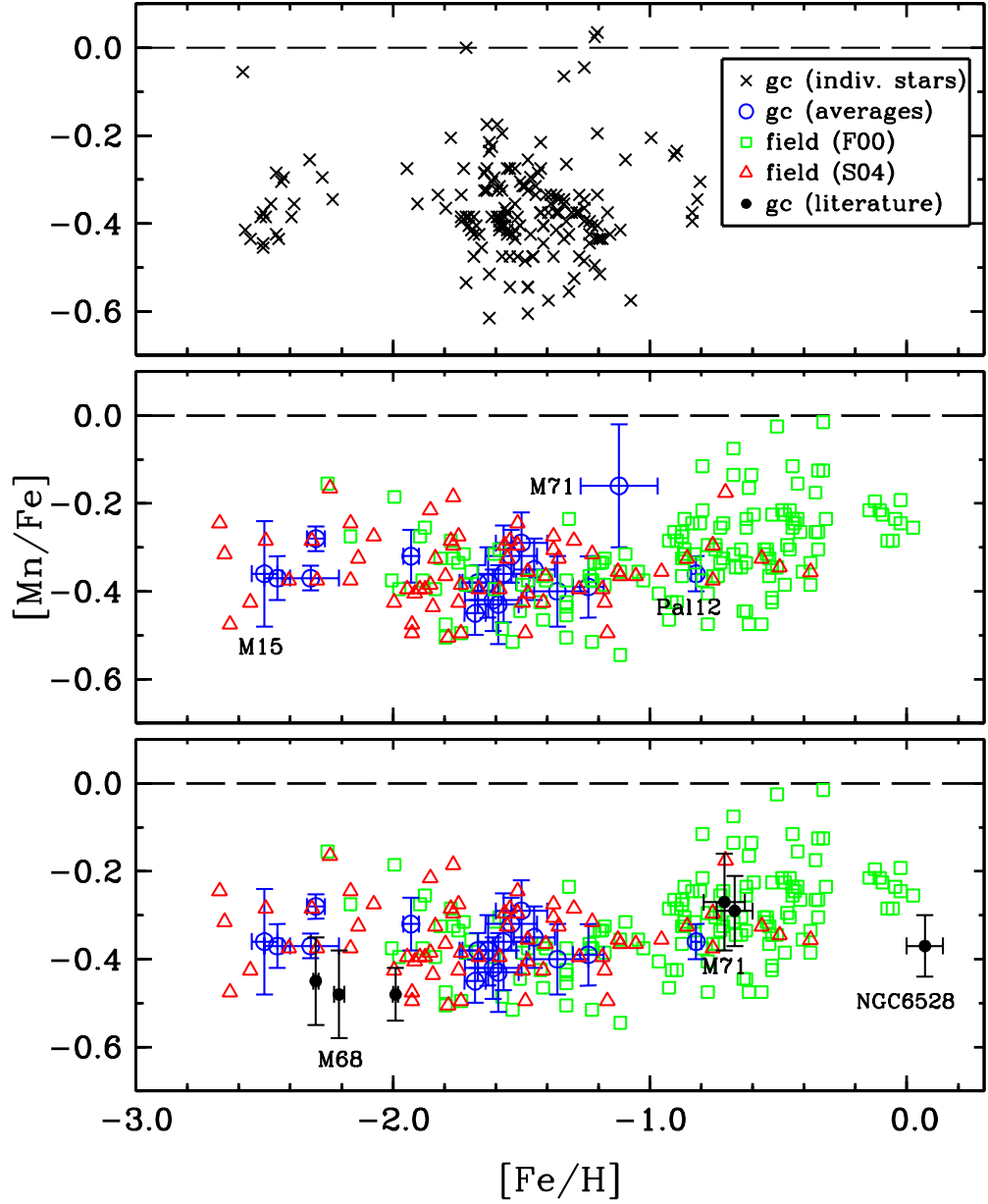}
\caption{Correlation of [Mn/Fe] with [Fe/H] for different stellar samples.  The top panel displays all of the abundances 
	 for the stars of the 19 globular cluster data sample.  The middle panel shows the average [Mn/Fe] and [Fe/H] values for 
         each globular cluster (with associated error bars).  Field star abundances are also shown in this panel; label F00 
	 indicates data from Fulbright (2000) and label S04 signifies data from Simmerer et al. (2004).
	 The points for M15 and Pal 12 are designated as they represent the extremes in metallicity for the halo globular clusters 
	 of this data set.  Moreover, M71 is denoted as its $<$[Mn/Fe]$>$ is not consistent with the other globular cluster data points.
	 The bottom panel presents globular cluster values from the literature.  Note that the literature Mn abundances
	 agree fairly well with those of the current study.  Also, the Mn abundance for M71 from the current data sample is set 
	 aside in favor of the value published by Ramirez \& Cohen (2002), as explained in \S 4.2.    
\label{f3}}
\end{figure}

The single-valued [Mn/Fe] ratio in the range $-$0.7$>$[Fe/H]$>$$-$2.7 may be described as a plateau (Figure ~\ref{f3}). 
Though the metallicity changes by roughly a factor of 100, $<$[Mn/Fe]$>$ does not vary in either globular cluster
or halo field stars.  In the specified range, the [Mn/Fe] ratio of (either stellar population) is not metallicity-dependent.
These data indicate that the contribution from stars that undergo core collapse SNe (i.e., medium to moderately high mass stars)
is uniform and does not change.  Furthermore, the data suggest that the initial mass function
(IMF) associated with these stars is essentially invariant.
As Thielemann et al. (1996\nocite{Thi96}) contend, in the range --1.0$\geq$[Fe/H]$\geq$--2.5, constant
abundance ratios of elements (like those of the Fe peak) should be expected as the core collapse SNe of the entire
mass range of progenitor stars occurs.  

Beyond [Fe/H]$\sim$ $-$1.0, there is an increase in the [Mn/Fe] scatter for the field star data points,  
and the relative Mn abundance rises steadily as solar metallicity is approached (Figure~\ref{f3}).  A possible
explanation for the increase in scatter is that in this metallicity regime stars of three
populations are present (halo, thin disk, and thick disk). Notably, Reddy et al. (2006\nocite{Red06}) 
present data that show no difference in the Mn abundance between the stars of the thin and thick disk (of the same metallicity).
The emergence of Type Ia SN events is likely responsible for the observable increase in the levels of Mn.  
This follows as the production of Mn occurs mostly in Type Ia SNe (e.g., Samland 1998\nocite{Sam98}; 
Iwamoto et al. 1999\nocite{Iwa99}).
 
It would be advantageous to use the [O/Mn] ratio in the examination of the evolution of very massive stars 
(the highest end of the IMF).  Virtually all synthesis of O occurs in massive stars.  
The full extent of the mass range of core collapse progenitors produces Fe (Thielemann et al.$\sigma$ = 0.10) 1996\nocite{Thi96}).  
Mn differs from Fe in that its manufacture occurs in a wide but limited portion of that mass range for Type II SNe.  
Consequently, the [O/Mn] ratio could provide constraints on the uppermost portion of the IMF.  
Unfortunately, as it pertains to this discussion, significant star-to-star variation of O abundance occurs in 
evolved stars of globular clusters with the diminution of 
O being due to the CNO and NeNa cycles of H burning (the proton-capture reactions; Denissenkov \& Weiss 2004\nocite{Den04}; 
Gratton et al. 2004\nocite{Gra04}).  As the bulk of the current study data is from globular cluster stars, 
little about nucleosynthesis in massive stars would be learned from [O/Mn] correlations.  
Work on this issue should be pursued with large field star 
samples that are extremely metal-poor ([Fe/H]$<-3.0$) or metal-rich ([Fe/H]$>-0.5$)~in nature.

Few recent theoretical reviews of elemental yields and abundances in the metallicity range of interest, 
$-$0.7$>$[Fe/H]$>$$-$2.7, have been published.  The comprehensive investigation by Timmes et al. 
(1995\nocite{Tim95}) examined the chemical evolution of 76 stable isotopes in this range
using the output from the Type II SN models of Woosley \& Weaver (1995\nocite{Woo95}).  Timmes et al. 
found excellent agreement between their calculations and the observational data for Cr and Ni.  
Although the trends for Mn, Sc, and V were well reproduced, the calculations of Timmes et al. predicted systematically lower 
abundance values for these elements than those found by observation.  The trend for Cu was fairly well duplicated, 
although the actual values for the calculated abundance were quite low in contrast to observational values.  The 
disagreement between theoretical calculations and observational results widens as the extremely low metallicity
regime is considered.  Limongi \& Chieffi (2005\nocite{Lim05}) compared their yields from zero metallicity core
collapse SNe to the extremely metal-poor star data of Cayrel et al. (2004\nocite{Cay04}).  
The observational data for the abundance ratios of the Fe-peak elements could not be simultaneously reproduced
by any of the models (regardless of the choice of mass cut).  This discrepancy 
encourages the continued development of theoretical calculations.

Further elucidation of the metal-rich regime is necessary, with special emphasis paid to bulge and disk clusters.  It must be
determined whether NGC 6528 is unique in its chemical evolution history 
(as suggested by McWilliam \& Rich 2004\nocite{McW04}) or, indeed, whether it is representative of all bulge clusters.  In addition, 
verification of the Mn abundance trend in the IR wavelength range and extension of this study to 
metal rich candidates is paramount.  It would also be valuable to re-investigate Fe-peak elements such as Co and Sc 
with large abundance uncertainties. 
 
\acknowledgments
We wish to thank to E. Carretta for many informative and insightful discussions.  We would like to acknowledge the 
data contributions of these investigators: J. Cohen, E. Friel, G. Gonzalez, F. Grundahl, J. W. Lee, 
M. Shetrone, and M. Zoccali.  We are deeply indebted to them.  We are grateful to the following agencies for providing funding 
support for this research: NASA, through Hubble Fellowship grant HST-HF-01151.01-A from the Space Telescope Science 
Institute, operated by AURA, Inc., under NASA contract NAS 5-26555 to I. I. I.; and the NSF, through grants 
AST 03-07495 to C. S. and AST 00-98453 to R. P. K.

\newpage

\newpage

\newpage

\newpage

\newpage

\newpage

\newpage

\newpage
\tablenum{1}
\tablecolumns{5}
\tablewidth{0pt}

\begin{deluxetable}{lclcr}

\tablecaption{LTG Observational Data}

\tablehead{
\colhead{Cluster}                          &                           
\colhead{Program Stars\tablenotemark{a}}   &
\colhead{Instrument\tablenotemark{b}}      &
\colhead{Reference\tablenotemark{c}}       &
\colhead{S/N Range\tablenotemark{d}}      \\
}
\startdata
NGC 5272 (M3)  & 20 & Keck(45)   & 1   & 40-150~~D\\
NGC 5904 (M5)  & 8  & Lick(30)   & 2   & 25-145~~A\\
               & 23 & Keck(45)   &     &          \\
NGC 6121 (M4)  & 20 & McD(30:60) & 3   & 40-170~~B\\
NGC 6205 (M13) & 17 & Lick(30)   & 4   & 65-180~~A\\
NGC 6254 (M10) & 12 & Lick(30)   & 5   & 60-180~~A\\
NGC 6341 (M92) & 4  & Keck(45)   & 6   & 60-85~~~A\\
NGC 6838 (M71) & 10 & Lick(30)   & 7   & 55-85~~~A\\
NGC 7006       & 6  & Keck(45)   & 8   & 55-95~~~B\\
NGC 7078 (M15) & 5  & Keck(45)   & 9   & 75-150~~C\\
               & 6  & Lick(50)   &     &          \\
Pal 5          & 4  & Keck(34)   & 10  & 60-100~~B\\
Halo Field Survey   & 130 & Lick(50)   & 11 & 40-240~~F\\
                    &  *  & Keck(45)   &    &          \\
                    &  *  & ESO (40)   &    &          \\
Halo Field Survey   & 86  & McD(60)    & 12 & $\simeq$100~~~~E\\
\enddata

\tablenotetext{a}{Note that the number of program stars does not always equal those found in the original paper.  In some
                  cases, the S/N ratio was too low in specified wavelength range to obtain a [Mn/Fe] ratio.}
\tablenotetext{b}{ESO (50): ESO 3.6m telescope-CASPEC spectrograph with R$\simeq$40,000;
                  Keck(34), Keck (45): Keck I 10.0m telescope-HIRES spectrograph with R$\simeq$34,000
                  or R$\simeq$45,000; Lick(30): Pre-1995 configuration of Lick 3.0m telescope-Hamilton echelle with
                  R$\simeq$30,000; Lick(50): Current configuration of Lick 3.0m telescope-Hamilton echelle with
                  R$\simeq$50,000; McD(30-60): McDonald 2.7m telescope-''2d-coud\'{e}'' with R$\simeq$30,000 or 
		  R$\simeq$60,000.}
\tablenotetext{c}{(1) Sneden et al. 2004. (2) Ivans et al. 2001. (3) Ivans et al. 1999. (4) Kraft et al. 1997. 
                  (5) Kraft et al. 1995. (6) Shetrone et al. 1998. (7) Sneden et al. 1994. (8) Kraft et al. 1998. 
                  (9) Sneden et al. 1997. (10) Smith et al. 2002. (11) Fulbright 2000. (12) Simmerer et al. 2004.}
\tablenotetext{d}{Approximate S/N at A: 6300 \AA; B: 6350 \AA; C: 6363 \AA; D: 6460 \AA; 
                  E: 4100 \AA; F: 5500 \AA.}

\end{deluxetable}

\newpage
\tablenum{2}
\tablecolumns{5}
\tablewidth{0pt}

\begin{deluxetable}{lclcl}

\tablecaption{Outside Source Cluster Observational Data}
\tablehead{
\colhead{Cluster}                          &                           
\colhead{Program Stars\tablenotemark{a}}   &        
\colhead{Instrument\tablenotemark{b}}      &
\colhead{Reference\tablenotemark{c}}       &
\colhead{S/N Range\tablenotemark{d}}      \\
}
\startdata
Cr~ 261        & 4  & CTIO(30)   & 1    & 75-100\\
NGC 288        & 13 & CTIO(30)   & 2    & 60-125~~B\\
NGC 362        & 12 & CTIO(30)   & 2    & 70-95~~~~B\\
NGC 3201       & 14 & CTIO(30)   & 3    & 40-70~~~~C\\
NGC 6287       & 2  & CTIO(30)   & 4    & $\simeq$95\\
NGC 6293       & 2  & CTIO(30)   & 4    & 95$\sim$100\\
NGC 6528       & 3  & VLT        & 5    & 30-40\\
NGC 6541       & 2  & CTIO(30)   & 4    & 95$\sim$135\\
NGC 6705 (M11) & 6  & CTIO(24)   & 6    & 85-130\\
               & 4  & APO(34)    &      &      \\
NGC 6752       & 4  & VLT(40:60) & 7, 8 & $\simeq$100~~~~B\\
Pal 12         & 4  & Keck(34)   & 9    & $>$100~~~~A\\
\enddata

\tablenotetext{a}{Note that the number of program stars does not always equal those found in the original paper.  In some
                  cases, the S/N ratio was too low in specified wavelength range to obtain a [Mn/Fe] ratio.}
\tablenotetext{b}{Keck(34): Keck I 10.0m telescope-HIRES spectrograph with R$\simeq$34,000;
                  CTIO(24), CTIO(30): CTIO 4.0m telescope-echelle spectrograph with R$\simeq$24,000 or R$\simeq$30,000;
                  VLT(40:60): VLT telescope-UVES spectrograph with R$\simeq$40,000 or R$\simeq$60,000;
                  APO(34): APO 3.5m telescope-echelle spectrograph with R$\simeq$34,000.}
\tablenotetext{c}{(1) Friel et al. 2003. (2) Shetrone \& Keane 2000. (3) Gonzalez \& Wallerstein 1998.
                  (4) Lee \& Carney 2002. (5) Zoccali et al. 2004. (6) Gonzalez \& Wallerstein 2000.
                  (7) Grundahl et al. 2002. (8) Yong et al. 2005. (9) Cohen 2004a.} 
\tablenotetext{d}{Approximate S/N at A: 5865 \AA; B: 6700 \AA; C: 7500 \AA}
\end{deluxetable}

\newpage
\tablenum{3}
\tablecolumns{8}
\tablewidth{0pt}

\begin{deluxetable}{lcccccccc}
\tablecaption{LTG Stellar Model Parameters and Individual [Fe/H] and [Mn/Fe] Values}
\tablehead{
\colhead{Association}                            &
\colhead{Star}                                   &
\colhead{T$_{eff}$}                              &
\colhead{log g}                                  &
\colhead{v$_t$}                                  &
\colhead{$[Fe/H]$}                               &
\colhead{$[Fe/H]$}                               &
\colhead{$[Mn/Fe]$}                             \\
\colhead{}                                       &
\colhead{}                                       &
\colhead{(K)}                                    &
\colhead{}                                       &
\colhead{(km s$^{-1}$)}                          &
\colhead{LIT}\tablenotemark{a}                   &
\colhead{}                                       &
\colhead{}                                       &
}
\startdata
NGC 5272 (M3) & B21&	4725&	1.65&	1.20&	$-$1.61&	$-$1.48 &	$-$0.61 \\
NGC 5272 (M3) & B23&	4800&	2.10&	1.40&	$-$1.70&	$-$1.63 &	$-$0.61 \\
NGC 5272 (M3) & B24&	4450&	1.00&	1.90&	$-$1.59&	$-$1.52 &	$-$0.48 \\
NGC 5272 (M3) & B33&	4550&	1.30&	1.65&	$-$1.55&	$-$1.48 &	$-$0.55 \\
NGC 5272 (M3) & B34&	3850&	0.00&	2.00&	$-$1.62&	$-$1.55 &	$-$0.55 \\
NGC 5272 (M3) & B11&	4400&	1.10&	1.50&	$-$1.56&	$-$1.59 &	$-$0.41 \\
NGC 5272 (M3) & B12&	4400&	1.10&	1.80&	$-$1.62&	$-$1.74 &	$-$0.34 \\
NGC 5272 (M3) & B13&	3900&	0.00&	2.05&	$-$1.54&	$-$1.58 &	$-$0.41 \\
NGC 5272 (M3) & B14&	4175&	0.70&	1.70&	$-$1.58&	$-$1.60 &	$-$0.39 \\
NGC 5272 (M3) & B15&	4350&	1.10&	1.50&	$-$1.67&	$-$1.67 &	$-$0.43 \\
NGC 5272 (M3) & B41&	4075&	0.40&	1.70&	$-$1.54&	$-$1.57 &	$-$0.41 \\
NGC 5272 (M3) & B42&	4100&	0.40&	1.70&	$-$1.56&	$-$1.56 &	$-$0.43 \\
NGC 5272 (M3) & B43&	4750&	1.40&	1.60&	$-$1.75&	$-$1.71 &	$-$0.39 \\
NGC 5272 (M3) & B44&	5050&	2.00&	1.50&	$-$1.71&	$-$1.71 &	$-$0.41 \\
NGC 5272 (M3) & B45&	5100&	2.40&	1.00&	$-$1.58&	$-$1.73 &	$-$0.28 \\
NGC 5272 (M3) & F24&	4600&	1.70&	1.20&	$-$1.54&	$-$1.51 &	$-$0.41 \\
NGC 5272 (M3) & I21&	4175&	0.70&	1.70&	$-$1.52&	$-$1.57 &	$-$0.39 \\
NGC 5272 (M3) & IV-101&	4200&	0.75&	1.70&	$-$1.50&	$-$1.46 &	$-$0.48 \\
NGC 5272 (M3) & IV-77&	4300&	0.85&	1.80&	$-$1.52&	$-$1.60 &	$-$0.33 \\
NGC 5272 (M3) & VZ1397&	3925&	0.10&	2.00&	$-$1.53&	$-$1.57 &	$-$0.37 \\
NGC 5904 (M5) &	G2 &	3900 &	-0.10 &	1.75 &	$-$1.33 &	$-$1.38 &	$-$0.48 \\
NGC 5904 (M5) &	I-14 &	4250 &	0.75 &	1.60 &	$-$1.34 &	$-$1.36 &	$-$0.38 \\
NGC 5904 (M5) &	I-2 &	4500 &	1.10 &	1.45 &	$-$1.31 &	$-$1.29 &	$-$0.38 \\
NGC 5904 (M5) &	I-20 &	4050 &	0.00 &	2.00 &	$-$1.44 &	$-$1.60 &	$-$0.18 \\
NGC 5904 (M5) &	I-50 &	4525 &	1.15 &	1.40 &	$-$1.33 &	$-$1.28 &	$-$0.48 \\
NGC 5904 (M5) &	I-55 &	4700 &	0.85 &	1.80 &	$-$1.47 &	$-$1.56 &	$-$0.38 \\
NGC 5904 (M5) &	I-58 &	4350 &	0.80 &	1.50 &	$-$1.27 &	$-$1.34 &	$-$0.35 \\
NGC 5904 (M5) &	I-61 &	4400 &	1.00 &	1.50 &	$-$1.32 &	$-$1.36 &	$-$0.35 \\
NGC 5904 (M5) &	I-68 &	4066 &	0.63 &	2.20 &	$-$1.44 &	$-$1.45 &	$-$0.58 \\ 
NGC 5904 (M5) &	I-71 &	4360 &	1.12 &	1.65 &	$-$1.32 &	$-$1.38 &	$-$0.37 \\
NGC 5904 (M5) &	II-50 &	4525 &	1.15 &	1.35 &	$-$1.24 &	$-$1.31 &	$-$0.38 \\
NGC 5904 (M5) &	II-59 &	4463 &	1.15 &	1.65 &	$-$1.33 &	$-$1.32 &	$-$0.56 \\
NGC 5904 (M5) &	II-74 &	4525 &	1.30 &	1.30 &	$-$1.17 &	$-$1.08 &	$-$0.58 \\
NGC 5904 (M5) &	II-85 &	4009 &	0.54 &	1.80 &	$-$1.30 &	$-$1.37 &	$-$0.31 \\
NGC 5904 (M5) &	III-122 &	4001 &	0.44 &	2.00 &	$-$1.26 &	$-$1.23 &	$-$0.54 \\
NGC 5904 (M5) &	III-18 &	4475 &	0.55 &	1.70 &	$-$1.43 &	$-$1.42 &	$-$0.45 \\
NGC 5904 (M5) &	III-3 &	4076 &	0.63 &	1.95 &	$-$1.31 &	$-$1.41 &	$-$0.41 \\
NGC 5904 (M5) &	III-36 &	4227 &	0.91 &	1.65 &	$-$1.28 &	$-$1.28 &	$-$0.47 \\
NGC 5904 (M5) &	III-52 &	4625 &	1.50 &	1.45 &	$-$1.38 &	$-$1.39 &	$-$0.36 \\
NGC 5904 (M5) &	III-53 &	4700 &	1.05 &	1.75 &	$-$1.52 &	$-$1.64 &	$-$0.33 \\
NGC 5904 (M5) &	III-59 &	4575 &	1.20 &	1.35 &	$-$1.30 &	$-$1.24 &	$-$0.45 \\
NGC 5904 (M5) &	III-78 &	4154 &	0.78 &	1.95 &	$-$1.32 &	$-$1.38 &	$-$0.45 \\
NGC 5904 (M5) &	IV-19 &	4125 &	0.50 &	1.70 &	$-$1.39 &	$-$1.32 &	$-$0.43 \\
NGC 5904 (M5) &	IV-26 &	4650 &	1.05 &	1.40 &	$-$1.41 &	$-$1.43 &	$-$0.38 \\
NGC 5904 (M5) &	IV-30 &	4625 &	1.00 &	1.75 &	$-$1.47 &	$-$1.47 &	$-$0.43 \\
NGC 5904 (M5) &	IV-34 &	4275 &	0.65 &	1.55 &	$-$1.28 &	$-$1.33 &	$-$0.40 \\
NGC 5904 (M5) &	IV-36 &	4575 &	1.50 &	1.35 &	$-$1.27 &	$-$1.28 &	$-$0.38 \\
NGC 5904 (M5) &	IV-4 &	4625 &	1.55 &	1.20 &	$-$1.24 &	$-$1.40 &	$-$0.38 \\
NGC 5904 (M5) &	IV-47 &	4110 &	0.50 &	1.85 &	$-$1.34 &	$-$1.30 &	$-$0.53 \\
NGC 5904 (M5) &	IV-59 &	4229 &	0.79 &	2.10 &	$-$1.40 &	$-$1.44 &	$-$0.45 \\
NGC 5904 (M5) &	IV-81 &	3945 &	0.00 &	1.90 &	$-$1.35 &	$-$1.37 &	$-$0.38 \\
NGC 6121 (M4) &	1408&	4525&	1.30&	1.70&	$-$1.18&	$-$1.23 &	$-$0.41 \\
NGC 6121 (M4) &	1411&	3950&	0.60&	1.65&	$-$1.21&	$-$1.26 &	$-$0.35 \\
NGC 6121 (M4) &	1514&	3875&	0.35&	1.95&	$-$1.23&	$-$1.34 &	$-$0.36 \\
NGC 6121 (M4) &	1701&	4625&	1.50&	1.65&	$-$1.20&	$-$1.20 &	$-$0.44 \\
NGC 6121 (M4) &	2206&	4325&	1.35&	1.55&	$-$1.18&	$-$1.21 &	$-$0.34 \\
NGC 6121 (M4) &	2208&	4350&	1.40&	1.70&	$-$1.12&	$-$1.17 &	$-$0.38 \\
NGC 6121 (M4) &	2307&	4075&	0.85&	1.45&	$-$1.20&	$-$1.28 &	$-$0.34 \\
NGC 6121 (M4) &	2406&	4100&	0.45&	2.45&	$-$1.22&	$-$1.26 &	$-$0.40 \\
NGC 6121 (M4) &	3207&	4700&	1.65&	1.70&	$-$1.18&	$-$1.21 &	$-$0.44 \\
NGC 6121 (M4) &	3209&	3975&	0.60&	1.75&	$-$1.22&	$-$1.26 &	$-$0.37 \\
NGC 6121 (M4) &	3215&	4775&	1.40&	1.85&	$-$1.17&	$-$1.24 &	$-$0.43 \\
NGC 6121 (M4) &	3413&	4175&	1.20&	1.65&	$-$1.18&	$-$1.33 &	$-$0.27 \\
NGC 6121 (M4) &	3612&	4250&	1.10&	1.45&	$-$1.20&	$-$1.21 &	$-$0.41 \\
NGC 6121 (M4) &	3624&	4225&	1.10&	1.45&	$-$1.16&	$-$1.24 &	$-$0.40 \\
NGC 6121 (M4) &	4201&	4450&	1.35&	1.85&	$-$1.19&	$-$1.19 &	$-$0.44 \\
NGC 6121 (M4) &	4302&	4775&	1.45&	1.80&	$-$1.18&	$-$1.22 &	$-$0.50 \\
NGC 6121 (M4) &	4511&	4150&	1.10&	1.55&	$-$1.16&	$-$1.16 &	$-$0.43 \\
NGC 6121 (M4) &	4513&	5250&	1.00&	1.65&	$-$1.20&	$-$1.43 &	$-$0.22 \\
NGC 6121 (M4) &	4611&	3725&	0.30&	1.70&	$-$1.16&	$-$1.20 &	$-$0.52 \\
NGC 6121 (M4) &	4613&	3750&	0.20&	1.65&	$-$1.19&	$-$1.26 &	$-$0.49 \\
NGC 6205 (M13) & L-629 &	3950 &	0.20 &	2.00 &	$-$1.68 &	$-$1.72 &	$-$0.36 \\
NGC 6205 (M13) & II-90 &	4000 &	0.30 &	2.00 &	$-$1.60 &	$-$1.65 &	$-$0.46 \\
NGC 6205 (M13) & II-67 &	3950 &	0.20 &	2.10 &	$-$1.58 &	$-$1.65 &	$-$0.37 \\
NGC 6205 (M13) & I-48  &	3920 &	0.30 &	2.00 &	$-$1.60 &	$-$1.66 &	$-$0.47 \\
NGC 6205 (M13) & L-598 &	3900 &	0.00 &	2.10 &	$-$1.64 &	$-$1.67 &	$-$0.38 \\
NGC 6205 (M13) & IV-22 &	4700 &	1.90 &	1.50 &	$-$1.56 &	$-$1.63 &	$-$0.37 \\
NGC 6205 (M13) & II-9  &	4700 &	1.70 &	1.50 &	$-$1.59 &	$-$1.63 &	$-$0.42 \\
NGC 6205 (M13) & II-28 &	4850 &	1.75 &	2.00 &	$-$1.68 &	$-$1.78 &	$-$0.33 \\
NGC 6205 (M13) & IV-25 &	4000 &	0.15 &	2.25 &	$-$1.61 &	$-$1.64 &	$-$0.37 \\
NGC 6205 (M13) & L-835 &	4090 &	0.55 &	1.90 &	$-$1.56 &	$-$1.63 &	$-$0.34 \\
NGC 6205 (M13) & I-54  &	4975 &	1.70 &	1.75 &	$-$1.71 &	$-$1.77 &	$-$0.41 \\
NGC 6205 (M13) & I-72  &	4850 &	1.90 &	1.45 &	$-$1.65 &	$-$1.75 &	$-$0.37 \\
NGC 6205 (M13) & II-1  &	4850 &	2.10 &	1.25 &	$-$1.58 &	$-$1.62 &	$-$0.44 \\
NGC 6205 (M13) & I-12  &	4600 &	1.50 &	1.60 &	$-$1.58 &	$-$1.66 &	$-$0.37 \\
NGC 6205 (M13) & IV-19 &	4650 &	1.50 &	1.60 &	$-$1.59 &	$-$1.64 &	$-$0.36 \\
NGC 6205 (M13) & II-41 &	4750 &	2.00 &	1.75 &	$-$1.51 &	$-$1.60 &	$-$0.36 \\
NGC 6205 (M13) &III-52 &        4335 &	1.00 &	2.00 &	$-$1.54 &	$-$1.72 &	$-$0.35 \\
NGC 6254 (M10) & A-I-2 & 3975   & 0.00 & 2.10 &	$-$1.47 &	$-$1.64 &	$-$0.33 \\
NGC 6254 (M10) & A-I-60 & 4400  & 1.10 & 1.60 &	$-$1.53 &	$-$1.48 &	$-$0.55 \\
NGC 6254 (M10) & A-I-61 & 4550  & 1.00 & 2.00 &	$-$1.69 &	$-$1.74 &	$-$0.40 \\
NGC 6254 (M10) & A-II-24 & 4050 & 0.10 & 2.00 &	$-$1.50 &	$-$1.55 &	$-$0.48 \\
NGC 6254 (M10) & A-III-16 & 4150 & 0.90 & 2.00 & $-$1.52 &	$-$1.62 &	$-$0.39 \\
NGC 6254 (M10) & A-III-21 & 4060 & 0.50 & 2.10 & $-$1.49 &	$-$1.64 &	$-$0.28 \\
NGC 6254 (M10) & A-III-5 & 4400 & 1.20 & 1.75 &	$-$1.36 &	$-$1.48 &	$-$0.26 \\
NGC 6254 (M10) & C & 4200 & 0.75 & 2.00 & $-$1.66 & $-$1.80 & $-$0.37 \\
NGC 6254 (M10) & D & 4200 & 1.05 & 2.00 & $-$1.50 & $-$1.59 & $-$0.40 \\
NGC 6254 (M10) & E & 4350 & 0.80 & 2.00 & $-$1.61 & $-$1.83 & $-$0.34 \\
NGC 6254 (M10) & H-I-15 & 4225 & 0.75 &	1.75 &	$-$1.52 & $-$1.59 &	$-$0.42 \\
NGC 6254 (M10) &H-I-367 & 4135 & 0.60 &	1.70 &	$-$1.54 & $-$1.68 &	$-$0.41 \\
NGC 6341 (M92) &	III-13 & 4180 &	0.10 &	2.15 &	$-$2.24 &	$-$2.39 &	$-$0.36 \\
NGC 6341 (M92) &	III-65 & 4260 &	0.30 &	1.80 &	$-$2.25 &	$-$2.46 &	$-$0.43 \\
NGC 6341 (M92) &	VII-122 & 4300 & 0.70 &	1.85 &	$-$2.32 &	$-$2.52 &	$-$0.39 \\
NGC 6341 (M92) &	VII-18 & 4220 &	0.20 &	2.00 &	$-$2.27 &	$-$2.43 &	$-$0.30 \\
NGC 6838 (M71) &	A4 &	4100 &	0.80 &	2.25 &	$-$0.78 &	$-$1.34 &	$-$0.07 \\
NGC 6838 (M71) &	I &	4300 &	1.00 &	2.00 &	$-$0.89 &	$-$1.26 &	$-$0.05 \\
NGC 6838 (M71) &	1-77 &	4100 &	0.95 &	2.00 &	$-$0.78 &	$-$1.22 &	0.03 \\
NGC 6838 (M71) &	1-45 &	4050 &	0.80 &	2.00 &	$-$0.76 & 	$-$1.21 &	$-$0.20 \\
NGC 6838 (M71) &	1-53 &	4300 &	1.40 &	2.00 &	$-$0.79 &	$-$1.21 &	0.04 \\
NGC 6838 (M71) &	1-113 &	3950 &	0.70 &	2.00 &	$-$0.85 & 	$-$1.12 &	$-$0.42 \\
NGC 6838 (M71) &	1-46 &	4000 &	0.80 &	2.15 &	$-$0.77 &	$-$1.10 &	$-$0.26 \\
NGC 6838 (M71) &	S &	4300 &	1.25 &	2.00 &	$-$0.72 &	$-$1.00 &	$-$0.21 \\
NGC 6838 (M71) &	1-21 &	4350 &	1.45 &	2.00 &	$-$0.73 &	$-$0.91 &	$-$0.25 \\
NGC 6838 (M71) &	A9 &	4200 &	1.20 &	2.00 &	$-$0.85 &	$-$0.90 &	$-$0.24 \\
NGC 7006 &	I-1 &	3900 &	0.10 &	2.25 &	$-$1.55 & $-$1.72 &	$-$0.39 \\
NGC 7006 &	II-103 & 4200 &	0.75 &	1.85 &	$-$1.55 & $-$1.58 &	$-$0.33 \\
NGC 7006 &	II-18 &	4300 &	0.90 &	1.85 &	$-$1.56 & $-$1.46 &	$-$0.48 \\
NGC 7006 &	II-46 &	4200 &	0.50 &	2.25 &	$-$1.60 & $-$1.54 &	$-$0.43 \\
NGC 7006 &	V19 &	4100 &	0.30 &	2.40 &	$-$1.62 & $-$1.69 &	$-$0.39 \\
NGC 7006 &	V54 &	4500 &	0.80 &	2.25 &	$-$1.65 & $-$1.72 &	$-$0.54 \\
NGC 7078 (M15) &	K341 &	4275 &	0.45 &	2.00 &	$-$2.35 &	$-$2.46 &	$-$0.29 \\
NGC 7078 (M15) &	K387 &	4400 &	0.65 &	1.85 &	$-$2.42 &	$-$2.51 &	$-$0.38 \\
NGC 7078 (M15) &	K969 &	4625 &	1.30 &	2.60 &	$-$2.42 &	$-$2.56 &	$-$0.44 \\
NGC 7078 (M15) &	K431 &	4375 &	0.50 &	2.30 &	$-$2.43 &	$-$2.50 &	$-$0.39 \\
NGC 7078 (M15) &	K146 &	4450 &	0.80 &	1.90 &	$-$2.46 &	$-$2.58 &	$-$0.42 \\
NGC 7078 (M15) &	K386 &	4200 &	0.15 &	1.85 &	$-$2.43 &	$-$2.51 &	$-$0.45 \\
NGC 7078 (M15) &	K583 &	4275 &	0.30 &	1.90 &	$-$2.40 &	$-$2.51 &	$-$0.46 \\
NGC 7078 (M15) &	K702 &	4325 &	0.25 &	1.90 &	$-$2.44 &	$-$2.45 &	$-$0.44 \\
NGC 7078 (M15) &	K462 &	4225 &	0.30 &	1.85 &	$-$2.45 &	$-$2.48 &	$-$0.36 \\
NGC 7078 (M15) &	K490 &	4350 &	0.60 &	1.65 &	$-$2.44 &	$-$2.59 &	$-$0.06 \\
NGC 7078 (M15) &	K634 &	4225 &	0.30 &	1.85 &	$-$2.38 &	$-$2.44 &	$-$0.31 \\
Pal 5 &	E &	4500 &	1.45 &	1.65 &	$-$1.39 &	$-$1.63 &	$-$0.22 \\
Pal 5 &	F &	4500 &	1.50 &	1.60 &	$-$1.33 &	$-$1.43 &	$-$0.38 \\
Pal 5 &	G &	4535 &	1.55 &	1.55 &	$-$1.31 &	$-$1.43 &	$-$0.28 \\
Pal 5 &	H &	4750 &	1.55 &	1.70 &	$-$1.32 &	$-$1.53 &	$-$0.28 \\
FIELD (JF) &	171	& 5275	& 4.1	& 1.05	& $-$1.00	& $-$0.91	& $-$0.27 \\
FIELD (JF) &	2413	& 5050	& 2.2	& 1.60	& $-$1.96	& $-$2.01	& $-$0.38 \\
FIELD (JF) &	3026	& 5950	& 3.9	& 1.40	& $-$1.32	& $-$1.32	& $-$0.24 \\
FIELD (JF) &	3086	& 5700	& 4.1	& 1.00	& $-$0.17	& $-$0.06	& $-$0.29 \\
FIELD (JF) &	5336	& 5250	& 4.4	& 0.90	& $-$0.98	& $-$0.83	& $-$0.32 \\
FIELD (JF) &	5445	& 5150	& 2.8	& 1.50	& $-$1.58	& $-$1.61	& $-$0.32 \\
FIELD (JF) &	5458	& 4450	& 1.4	& 1.55	& $-$1.04	& $-$0.87	& $-$0.33 \\
FIELD (JF) &	6710	& 4625	& 1.2	& 1.95	& $-$1.83	& $-$1.90	& $-$0.28 \\
FIELD (JF) &	7217	& 5550	& 4.2	& 0.70	& $-$0.48	& $-$0.41	& $-$0.36 \\
FIELD (JF) &	10140	& 5425	& 4.1	& 0.85	& $-$1.14	& $-$0.97	& $-$0.41 \\
FIELD (JF) &	10449	& 5650	& 4.4	& 1.00	& $-$0.98	& $-$0.90	& $-$0.43 \\
FIELD (JF) &	11349	& 5375	& 4.3	& 0.80	& $-$0.29	& $-$0.15	& $-$0.22 \\
FIELD (JF) &	12306	& 5650	& 4.1	& 1.05	& $-$0.63	& $-$0.51	& $-$0.35 \\
FIELD (JF) &	13366	& 5700	& 4.2	& 0.95	& $-$0.77	& $-$0.72	& $-$0.33 \\
FIELD (JF) &	14086	& 5075	& 3.6	& 1.10	& $-$0.71	& $-$0.68	& $-$0.32 \\
FIELD (JF) &	15394	& 5150	& 3.4	& 1.00	& $-$0.30	& $-$0.08	& $-$0.29 \\
FIELD (JF) &	16214	& 4825	& 2.0	& 1.45	& $-$1.74	& $-$1.72	& $-$0.39 \\
FIELD (JF) &	17085	& 6500	& 4.2	& 1.70	& $-$0.22	& $-$0.13	& $-$0.20 \\
FIELD (JF) &	17147	& 5800	& 4.3	& 1.10	& $-$0.91	& $-$0.83	& $-$0.34 \\
FIELD (JF) &	17666	& 5050	& 4.5	& 0.60	& $-$1.10	& $-$0.93	& $-$0.47 \\
FIELD (JF) &	18235	& 4950	& 3.2	& 0.90	& $-$0.72	& $-$0.63	& $-$0.45 \\
FIELD (JF) &	18915	& 4700	& 4.8	& 1.35	& $-$1.85	& $-$1.77	& $-$0.33 \\
FIELD (JF) &	18995	& 5575	& 2.2	& 2.05	& $-$1.26	& $-$1.24	& $-$0.37 \\
FIELD (JF) &	19007	& 5150	& 4.5	& 1.20	& $-$0.62	& $-$0.51	& $-$0.03 \\
FIELD (JF) &	19378	& 4500	& 1.2	& 1.70	& $-$1.73	& $-$1.75	& $-$0.34 \\
FIELD (JF) &	21000	& 6200	& 4.1	& 1.40	& $-$0.16	& $-$0.11	& $-$0.22 \\
FIELD (JF) &	21586	& 4850	& 4.1	& 0.25	& $-$0.91	& $-$0.67	& $-$0.35 \\
FIELD (JF) &	21609	& 5200	& 3.8	& 1.55	& $-$1.76	& $-$1.70	& $-$0.36 \\
FIELD (JF) &	21648	& 4300	& 0.4	& 1.70	& $-$1.88	& $-$1.84	& $-$0.40 \\
FIELD (JF) &	21767	& 5650	& 4.5	& 0.70	& $-$0.44	& $-$0.35	& $-$0.31 \\
FIELD (JF) &	22246	& 5200	& 4.5	& 1.20	& $-$0.38	& $-$0.33	& $-$0.13 \\
FIELD (JF) &	22632	& 5825	& 4.3	& 1.35	& $-$1.59	& $-$1.61	& $-$0.40 \\
FIELD (JF) &	26688	& 6500	& 4.1	& 1.50	& $-$0.60	& $-$0.61	& $-$0.14 \\
FIELD (JF) &	27654	& 4550	& 2.1	& 1.50	& $-$0.94	& $-$0.87	& $-$0.32 \\
FIELD (JF) &	28188	& 6175	& 4.6	& 1.25	& $-$0.62	& $-$0.63	& $-$0.17 \\
FIELD (JF) &	30668	& 5150	& 3.1	& 1.05	& $-$1.50	& $-$1.54	& $-$0.52 \\
FIELD (JF) &	30990	& 5825	& 4.0	& 1.30	& $-$0.89	& $-$0.93	& $-$0.29 \\
FIELD (JF) &	31188	& 5750	& 4.1	& 1.65	& $-$0.80	& $-$0.63	& $-$0.24 \\
FIELD (JF) &	31639	& 5300	& 4.3	& 0.60	& $-$0.62	& $-$0.47	& $-$0.30 \\
FIELD (JF) &	32308	& 5175	& 4.1	& 1.00	& $-$0.64	& $-$0.42	& $-$0.27 \\
FIELD (JF) &	33582	& 5725	& 4.3	& 1.25	& $-$0.74	& $-$0.80	& $-$0.12 \\
FIELD (JF) &	34146	& 6300	& 4.2	& 1.95	& $-$0.40	& $-$0.43	& $-$0.16 \\
FIELD (JF) &	34548	& 6250	& 4.5	& 1.40	& $-$0.46	& $-$0.45	& $-$0.12 \\
FIELD (JF) &	36491	& 5800	& 4.4	& 1.10	& $-$0.93	& $-$0.87	& $-$0.31 \\
FIELD (JF) &	36849	& 5850	& 4.1	& 1.10	& $-$0.88	& $-$0.88	& $-$0.27 \\
FIELD (JF) &	38541	& 5300	& 4.7	& 0.85	& $-$1.79	& $-$1.74	& $-$0.50 \\
FIELD (JF) &	38621	& 4700	& 1.7	& 2.25	& $-$1.81	& $-$1.88	& $-$0.26 \\
FIELD (JF) &	38625	& 5200	& 4.4	& 0.30	& $-$0.86	& $-$0.73	& $-$0.34 \\
FIELD (JF) &	40068	& 5225	& 3.0	& 1.35	& $-$2.05	& $-$1.98	& $-$0.40 \\
FIELD (JF) &	44075	& 5900	& 4.2	& 1.25	& $-$0.91	& $-$0.90	& $-$0.29 \\
FIELD (JF) &	44116	& 6275	& 4.1	& 1.45	& $-$0.58	& $-$0.53	& $-$0.23 \\
FIELD (JF) &	44716	& 5000	& 2.1	& 1.70	& $-$1.08	& $-$1.10	& $-$0.32 \\
FIELD (JF) &	44919	& 6350	& 3.8	& 1.80	& $-$0.65	& $-$0.68	& $-$0.08 \\
FIELD (JF) &	47139	& 4600	& 1.3	& 1.80	& $-$1.46	& $-$1.48	& $-$0.38 \\
FIELD (JF) &	47640	& 6600	& 4.4	& 1.50	& $-$0.08	& $-$0.05	& $-$0.24 \\
FIELD (JF) &	48146	& 6200	& 4.6	& 1.05	& $-$0.05	& $-$0.02	& $-$0.19 \\
FIELD (JF) &	49371	& 4950	& 2.3	& 1.75	& $-$1.95	& $-$1.89	& $-$0.38 \\
FIELD (JF) &	50139	& 5600	& 4.3	& 0.35	& $-$0.68	& $-$0.56	& $-$0.31 \\
FIELD (JF) &	54858	& 5250	& 2.0	& 2.15	& $-$1.17	& $-$1.20	& $-$0.34 \\
FIELD (JF) &	57265	& 5875	& 4.0	& 1.50	& $-$1.10	& $-$1.05	& $-$0.36 \\
FIELD (JF) &	57850	& 4375	& 0.8	& 2.75	& $-$1.78	& $-$1.78	& $-$0.29 \\
FIELD (JF) &	58229	& 5875	& 4.1	& 1.25	& $-$0.94	& $-$0.92	& $-$0.43 \\
FIELD (JF) &	58357	& 5050	& 3.4	& 1.20	& $-$0.65	& $-$0.72	& $-$0.25 \\
FIELD (JF) &	59239	& 5125	& 2.1	& 1.55	& $-$1.49	& $-$1.50	& $-$0.39 \\
FIELD (JF) &	59330	& 5750	& 4.1	& 1.25	& $-$0.75	& $-$0.73	& $-$0.26 \\
FIELD (JF) &	59750	& 6200	& 4.4	& 1.10	& $-$0.78	& $-$0.64	& $-$0.45 \\
FIELD (JF) &	60551	& 5725	& 4.4	& 1.05	& $-$0.86	& $-$0.87	& $-$0.24 \\
FIELD (JF) &	62747	& 4285	& 2.2	& 1.45	& $-$1.54	& $-$1.51	& $-$0.45 \\
FIELD (JF) &	62882	& 5600	& 3.7	& 0.04	& $-$1.26	& $-$1.12	& $-$0.55 \\
FIELD (JF) &	63970	& 6075	& 4.4	& 1.00	& $-$0.09	& 0.03	& $-$0.26 \\
FIELD (JF) &	64115	& 4650	& 2.4	& 1.10	& $-$0.74	& $-$0.62	& $-$0.48 \\
FIELD (JF) &	64426	& 5800	& 4.1	& 1.25	& $-$0.82	& $-$0.78	& $-$0.30 \\
FIELD (JF) &	65268	& 6250	& 4.1	& 1.50	& $-$0.67	& $-$0.60	& $-$0.23 \\
FIELD (JF) &	66246	& 4400	& 1.0	& 2.55	& $-$1.91	& $-$2.00	& $-$0.19 \\
FIELD (JF) &	66509	& 5350	& 4.2	& 0.60	& $-$0.68	& $-$0.53	& $-$0.43 \\
FIELD (JF) &	66665	& 5500	& 3.8	& 1.05	& $-$0.97	& $-$0.78	& $-$0.48 \\
FIELD (JF) &	66815	& 5875	& 4.5	& 0.95	& $-$0.64	& $-$0.63	& $-$0.26 \\
FIELD (JF) &	68796	& 5725	& 4.5	& 0.90	& $-$0.52	& $-$0.43	& $-$0.24 \\
FIELD (JF) &	68807	& 4575	& 1.1	& 1.90	& $-$1.83	& $-$1.82	& $-$0.33 \\
FIELD (JF) &	70681	& 5450	& 4.5	& 0.80	& $-$1.25	& $-$1.23	& $-$0.38 \\
FIELD (JF) &	71886	& 6400	& 4.1	& 1.50	& $-$0.40	& $-$0.32	& $-$0.24 \\
FIELD (JF) &	71887	& 6100	& 4.3	& 1.20	& $-$0.49	& $-$0.45	& $-$0.26 \\
FIELD (JF) &	71939	& 6300	& 4.4	& 1.50	& $-$0.37	& $-$0.36	& $-$0.18 \\
FIELD (JF) &	73385	& 5575	& 3.6	& 1.35	& $-$1.59	& $-$1.59	& $-$0.31 \\
FIELD (JF) &	73960	& 4500	& 1.4	& 2.10	& $-$1.37	& $-$1.33	& $-$0.38 \\
FIELD (JF) &	74033	& 5675	& 4.1	& 1.05	& $-$0.78	& $-$0.80	& $-$0.38 \\
FIELD (JF) &	74067	& 5575	& 4.3	& 1.10	& $-$0.90	& $-$0.88	& $-$0.28 \\
FIELD (JF) &	74079	& 5825	& 4.0	& 1.30	& $-$0.83	& $-$0.75	& $-$0.31 \\
FIELD (JF) &	74234	& 4750	& 4.5	& 0.70	& $-$1.51	& $-$1.33	& $-$0.51 \\
FIELD (JF) &	74235	& 4850	& 4.5	& 0.70	& $-$1.57	& $-$1.42	& $-$0.47 \\
FIELD (JF) &	80837	& 5800	& 4.1	& 1.15	& $-$0.83	& $-$0.74	& $-$0.37 \\
FIELD (JF) &	81170	& 5175	& 4.7	& 0.30	& $-$1.26	& $-$1.23	& $-$0.38 \\
FIELD (JF) &	81461	& 5600	& 4.1	& 1.20	& $-$0.65	& $-$0.46	& $-$0.31 \\
FIELD (JF) &	85007	& 5900	& 4.2	& 1.20	& $-$0.50	& $-$0.38	& $-$0.34 \\
FIELD (JF) &	85378	& 5625	& 4.0	& 1.10	& $-$0.64	& $-$0.53	& $-$0.35 \\
FIELD (JF) &	85757	& 5450	& 3.8	& 1.05	& $-$0.76	& $-$0.65	& $-$0.35 \\
FIELD (JF) &	86013	& 5750	& 4.4	& 1.15	& $-$0.82	& $-$0.84	& $-$0.25 \\
FIELD (JF) &	86431	& 5675	& 4.1	& 1.15	& $-$0.64	& $-$0.54	& $-$0.37 \\
FIELD (JF) &	88010	& 5200	& 4.0	& 0.70	& $-$1.49	& $-$1.41	& $-$0.38 \\
FIELD (JF) &	88039	& 5700	& 4.0	& 1.30	& $-$0.96	& $-$0.88	& $-$0.33 \\
FIELD (JF) &	91058	& 6025	& 4.1	& 1.40	& $-$0.54	& $-$0.49	& $-$0.22 \\
FIELD (JF) &	92167	& 4575	& 2.4	& 1.40	& $-$1.47	& $-$1.80	& $-$0.51 \\
FIELD (JF) &	92532	& 5825	& 4.3	& 1.00	& $-$0.56	& $-$0.44	& $-$0.25 \\
FIELD (JF) &	92781	& 5650	& 4.2	& 0.95	& $-$0.75	& $-$0.57	& $-$0.48 \\
FIELD (JF) &	94449	& 5625	& 3.7	& 1.15	& $-$1.26	& $-$1.22	& $-$0.34 \\
FIELD (JF) &	96185	& 5700	& 4.1	& 1.00	& $-$0.58	& $-$0.53	& $-$0.35 \\
FIELD (JF) &	97023	& 5800	& 3.8	& 1.30	& $-$0.48	& $-$0.35	& $-$0.27 \\
FIELD (JF) &	97468	& 4450	& 1.1	& 1.90	& $-$1.71	& $-$1.73	& $-$0.32 \\
FIELD (JF) &	98020	& 5325	& 4.6	& 1.10	& $-$1.67	& $-$1.58	& $-$0.49 \\
FIELD (JF) &	98532	& 5550	& 3.6	& 1.30	& $-$1.23	& $-$1.18	& $-$0.33 \\
FIELD (JF) &	99423	& 5650	& 3.8	& 1.30	& $-$1.50	& $-$1.43	& $-$0.43 \\
FIELD (JF) &	99938	& 5650	& 4.0	& 1.20	& $-$0.74	& $-$0.65	& $-$0.31 \\
FIELD (JF) &	100568	& 5650	& 4.4	& 1.10	& $-$1.17	& $-$1.12	& $-$0.36 \\
FIELD (JF) &	100792	& 5875	& 4.2	& 1.40	& $-$1.23	& $-$1.19	& $-$0.34 \\
FIELD (JF) &	101346	& 6000	& 3.9	& 1.40	& $-$0.65	& $-$0.68	& $-$0.14 \\
FIELD (JF) &	101382	& 5125	& 4.0	& 0.40	& $-$0.66	& $-$0.38	& $-$0.39 \\
FIELD (JF) &	103269	& 5300	& 4.6	& 0.85	& $-$1.81	& $-$1.80	& $-$0.48 \\
FIELD (JF) &	104659	& 5825	& 4.3	& 1.00	& $-$1.12	& $-$1.03	& $-$0.38 \\
FIELD (JF) &	104660	& 5500	& 3.9	& 1.15	& $-$0.96	& $-$0.78	& $-$0.41 \\
FIELD (JF) &	105888	& 5700	& 4.3	& 1.00	& $-$0.75	& $-$0.63	& $-$0.34 \\
FIELD (JF) &	107975	& 6275	& 3.9	& 1.50	& $-$0.64	& $-$0.54	& $-$0.33 \\
FIELD (JF) &	109067	& 5300	& 4.3	& 0.85	& $-$0.97	& $-$0.88	& $-$0.37 \\
FIELD (JF) &	109390	& 4800	& 2.2	& 1.50	& $-$1.34	& $-$1.33	& $-$0.43 \\
FIELD (JF) &	112796	& 4525	& 1.0	& 2.85	& $-$2.25	& $-$2.26	& $-$0.16 \\
FIELD (JF) &	114962	& 5825	& 4.3	& 1.40	& $-$1.54	& $-$1.33	& $-$0.44 \\
FIELD (JF) &	115610	& 4800	& 4.1	& 1.20	& $-$0.63	& $-$0.35	& $-$0.13 \\
FIELD (JF) &	115949	& 4500	& 0.9	& 2.75	& $-$2.19	& $-$2.17	& $-$0.28 \\
FIELD (JF) &	116082	& 6275	& 3.7	& 1.60	& $-$0.82	& $-$0.80	& $-$0.22 \\
FIELD (JF) &	117029	& 5425	& 3.8	& 1.05	& $-$0.81	& $-$0.75	& $-$0.30 \\
FIELD (JF) &	117041	& 5300	& 4.2	& 0.90	& $-$0.88	& $-$0.81	& $-$0.25 \\
FIELD (JS) &	B-010306	& 5550	& 4.19	& 1.50	&  $-$1.13	& $-$1.10	& $-$0.30 \\
FIELD (JS) &	B-012582	& 5148	& 2.86	& 1.20	& $-$2.21	& $-$2.26	& $-$0.41 \\
FIELD (JS) &	B+191185	& 5500	& 4.19	& 1.10	& $-$1.09	& $-$1.09	& $-$0.31 \\
FIELD (JS) &	B+521601	& 4911	& 2.10	& 2.05	& $-$1.40	& $-$1.49	& $-$0.33 \\
FIELD (JS) &	G005-001	& 5500	& 4.32	& 0.80	& $-$1.24	& $-$1.18	& $-$0.31 \\
FIELD (JS) &	G009-036	& 5625	& 4.57	& 0.65	& $-$1.17	& $-$1.17	& $-$0.44 \\
FIELD (JS) &	G017-025	& 4966	& 4.26	& 0.80	& $-$1.54	& $-$1.37	& $-$0.42 \\
FIELD (JS) &	G023-014	& 5025	& 3.00	& 1.30	& $-$1.64	& $-$1.57	& $-$0.44 \\
FIELD (JS) &	G028-043	& 5061	& 4.50	& 0.80	& $-$1.64	& $-$1.58	& $-$0.42 \\
FIELD (JS) &	G029-025	& 5225	& 4.28	& 0.80	& $-$1.09	& $-$0.91	& $-$0.44 \\
FIELD (JS) &	G040-008	& 5200	& 4.08	& 0.50	& $-$0.97	& $-$0.80	& $-$0.37 \\
FIELD (JS) &	G058-025	& 6001	& 4.21	& 1.05	& $-$1.40	& $-$1.49	& $-$0.36 \\
FIELD (JS) &	G059-001	& 5922	& 3.98	& 0.40	& $-$0.95	& $-$0.76	& $-$0.38 \\
FIELD (JS) &	G063-046	& 5705	& 4.25	& 1.30	& $-$0.90	& $-$0.85	& $-$0.28 \\
FIELD (JS) &	G068-003	& 4975	& 3.50	& 0.95	& $-$0.76	& $-$0.65	& $-$0.28 \\
FIELD (JS) &	G074-005	& 5668	& 4.24	& 1.50	& $-$1.05	& $-$1.03	& $-$0.28 \\
FIELD (JS) &	G090-025	& 5303	& 4.46	& 1.20	& $-$1.78	& $-$1.83	& $-$0.53 \\
FIELD (JS) &	G095-57A	& 4965	& 4.40	& 0.90	& $-$1.22	& $-$1.08	& $-$0.37 \\
FIELD (JS) &	G095-57B	& 4800	& 4.57	& 0.60	& $-$1.06	& $-$1.03	& $-$0.30 \\
FIELD (JS) &	G102-020	& 5254	& 4.44	& 0.90	& $-$1.25	& $-$1.23	& $-$0.32 \\
FIELD (JS) &	G102-027	& 5600	& 3.75	& 1.05	& $-$0.59	& $-$0.50	& $-$0.35 \\
FIELD (JS) &	G113-022	& 5525	& 4.25	& 1.10	& $-$1.18	& $-$1.19	& $-$0.40 \\
FIELD (JS) &	G122-051	& 4864	& 4.51	& 1.40	& $-$1.43	& $-$1.42	& $-$0.43 \\
FIELD (JS) &	G123-009	& 5487	& 4.75	& 1.50	& $-$1.25	& $-$1.30	& $-$0.29 \\
FIELD (JS) &	G126-036	 & 5487	& 4.50	& 0.60	& $-$1.06	& $-$0.96	& $-$0.36 \\
FIELD (JS) &	G126-062	& 5941	& 3.98	& 2.00	& $-$1.59	& $-$1.75	& $-$0.28 \\
FIELD (JS) &	G140-046	& 4980	& 4.42	& 0.70	& $-$1.30	& $-$1.18	& $-$0.43 \\
FIELD (JS) &	G153-021	& 5700	& 4.36	& 1.40	& $-$0.70	& $-$0.71	& $-$0.18 \\
FIELD (JS) &	G176-053	& 5593	& 4.50	& 1.20	& $-$1.34	& $-$1.41	& $-$0.37 \\
FIELD (JS) &	G179-022	& 5082	& 3.20	& 1.20	& $-$1.35	& $-$1.28	& $-$0.40 \\
FIELD (JS) &	G180-024	& 6059	& 4.09	& 0.50	& $-$1.34	& $-$1.38	& $-$0.28 \\
FIELD (JS) &	G188-022	& 5827	& 4.27	& 1.20	& $-$1.52	& $-$1.48	& $-$0.36 \\
FIELD (JS) &	G191-055	& 5770	& 4.50	& 1.00	& $-$1.63	& $-$1.77	& $-$0.19 \\
FIELD (JS) &	G192-043	& 6085	& 4.50	& 1.50	& $-$1.50	& $-$1.56	& $-$0.33 \\
FIELD (JS) &	G221-007	& 5016	& 3.37	& 0.90	& $-$0.98	& $-$0.86	& $-$0.33 \\
FIELD (JS) &	2665	& 4990	& 2.34	& 2.00	& $-$1.99	& $-$2.17	& $-$0.38 \\
FIELD (JS) &	3008	& 4250	& 0.25	& 2.60	& $-$2.08	& $-$2.14	& $-$0.33 \\
FIELD (JS) &	6755	& 5105	& 2.93	& 2.50	& $-$1.68	& $-$1.78	& $-$0.29 \\
FIELD (JS) &	8724	& 4535	& 1.40	& 1.40	& $-$1.91	& $-$1.79	& $-$0.51 \\
FIELD (JS) &	21581	& 4870	& 2.27	& 1.40	& $-$1.71	& $-$1.75	& $-$0.43 \\
FIELD (JS) &	23798	& 4450	& 1.06	& 2.50	& $-$2.26	& $-$2.32	& $-$0.29 \\
FIELD (JS) &	25329	& 4842	& 4.66	& 0.60	& $-$1.67	& $-$1.67	& $-$0.40 \\
FIELD (JS) &	25532	& 5396	& 2.00	& 1.20	& $-$1.34	& $-$1.17	& $-$0.50 \\
FIELD (JS) &	26297	& 4322	& 1.11	& 1.80	& $-$1.98	& $-$1.92	& $-$0.41 \\
FIELD (JS) &	29574	& 4250	& 0.80	& 2.20	& $-$2.00	& $-$2.00	& $-$0.43 \\
FIELD (JS) &	37828	& 4350	& 1.50	& 1.85	& $-$1.62	& $-$1.59	& $-$0.40 \\
FIELD (JS) &	44007	& 4851	& 2.00	& 2.00	& $-$1.72	& $-$1.74	& $-$0.39 \\
FIELD (JS) &	63791	& 4675	& 2.00	& 2.00	& $-$1.90	& $-$1.86	& $-$0.39 \\
FIELD (JS) &	74462	& 4700	& 2.00	& 1.90	& $-$1.52	& $-$1.55	& $-$0.29 \\
FIELD (JS) &	82590	& 6005	& 2.75	& 3.00	& $-$1.50	& $-$1.57	& $-$0.30 \\
FIELD (JS) &	85773	& 4268	& 0.50	& 2.00	& $-$2.62	& $-$2.50	& $-$0.29 \\
FIELD (JS) &	101063	& 5150	& 3.25	& 1.70	& $-$1.33	& $-$1.38	& $-$0.31 \\
FIELD (JS) &	103036	& 4200	& 0.25	& 3.00	& $-$2.04	& $-$1.93	& $-$0.48 \\
FIELD (JS) &	103545	& 4666	& 1.64	& 2.00	& $-$2.45	& $-$2.41	& $-$0.38 \\
FIELD (JS) &	105546	& 5190	& 2.49	& 1.60	& $-$1.48	& $-$1.52	& $-$0.25 \\
FIELD (JS) &	105755	& 5701	& 3.82	& 1.20	& $-$0.83	& $-$0.76	& $-$0.30 \\
FIELD (JS) &	106516	& 6166	& 4.21	& 1.10	& $-$0.81	& $-$0.76	& $-$0.38 \\
FIELD (JS) &	108317	& 5234	& 2.68	& 2.00	& $-$2.18	& $-$2.25	& $-$0.17 \\
FIELD (JS) &	110184	& 4250	& 0.50	& 2.50	& $-$2.72	& $-$2.66	& $-$0.32 \\
FIELD (JS) &	121135	& 4934	& 1.91	& 1.60	& $-$1.54	& $-$1.49	& $-$0.50 \\
FIELD (JS) &	122563	& 4572	& 1.36	& 2.90	& $-$2.72	& $-$2.68	& $-$0.25 \\
FIELD (JS) &	122956	& 4508	& 1.55	& 1.60	& $-$1.95	& $-$1.85	& $-$0.44 \\
FIELD (JS) &	124358	& 4688	& 1.57	& 2.10	& $-$1.91	& $-$1.88	& $-$0.40 \\
FIELD (JS) &	132475	& 5425	& 3.56	& 2.30	& $-$1.86	& $-$1.80	& $-$0.37 \\
FIELD (JS) &	135148	& 4183	& 0.25	& 2.90	& $-$2.17	& $-$2.17	& $-$0.25 \\
FIELD (JS) &	141531	& 4356	& 1.14	& 2.20	& $-$1.79	& $-$1.84	& $-$0.33 \\
FIELD (JS) &	165195	& 4237	& 0.78	& 2.30	& $-$2.60	& $-$2.56	& $-$0.43 \\
FIELD (JS) &	166161	& 5350	& 2.56	& 2.25	& $-$1.23	& $-$1.36	& $-$0.33 \\
FIELD (JS) &	171496	& 4952	& 2.37	& 1.40	& $-$0.67	& $-$0.57	& $-$0.33 \\
FIELD (JS) &	184266	& 6000	& 2.74	& 3.00	& $-$1.43	& $-$1.52	& $-$0.30 \\
FIELD (JS) &	186478	& 4598	& 1.43	& 2.00	& $-$2.56	& $-$2.64	& $-$0.48 \\
FIELD (JS) &	187111	& 4271	& 1.05	& 1.90	& $-$1.97	& $-$1.90	& $-$0.40 \\
FIELD (JS) &	188510	& 5564	& 4.51	& 1.00	& $-$1.32	& $-$1.50	& $-$0.43 \\
FIELD (JS) &	193901	& 5750	& 4.46	& 1.50	& $-$1.08	& $-$1.13	& $-$0.36 \\
FIELD (JS) &	194598	& 6044	& 4.19	& 1.00	& $-$1.08	& $-$1.12	& $-$0.37 \\
FIELD (JS) &	201891	& 5909	& 4.19	& 1.00	& $-$1.09	& $-$1.06	& $-$0.37 \\
FIELD (JS) &	204543	& 4672	& 1.49	& 2.00	& $-$1.87	& $-$1.95	& $-$0.40 \\
FIELD (JS) &	206739	& 4647	& 1.78	& 1.90	& $-$1.72	& $-$1.77	& $-$0.30 \\
FIELD (JS) &	210295	& 4750	& 2.50	& 1.55	& $-$1.46	& $-$1.48	& $-$0.41 \\
FIELD (JS) &	214362	& 5727	& 2.62	& 2.00	& $-$1.87	& $-$1.93	& $-$0.50 \\
FIELD (JS) &	218857	& 5103	& 2.44	& 1.90	& $-$1.90	& $-$2.08	& $-$0.28 \\
FIELD (JS) &	221170	& 4410	& 1.09	& 1.70	& $-$2.35	& $-$2.30	& $-$0.38 \\
FIELD (JS) &	232078	& 3875	& 0.50	& 2.10	& $-$1.69	& $-$1.74	& $-$0.50 \\
FIELD (JS) &	233666	& 5157	& 2.00	& 1.70	& $-$1.79	& $-$1.86	& $-$0.22 \\
\enddata
\tablecomments{The complete version of this table is in the electronic
edition of the Journal.  The printed edition contains only a sample.}
\tablenotetext{a}{As discussed in text, literature values of [Fe/H] are provided.}
\tablecomments{Field stars from the Fulbright (2000) survey (labeled 
               as JF) have Hipparcos identifications. Similarly, target stars of the 
               Simmerer et al. (2004) survey (designated as JS) have Henry Draper identifications unless
               otherwise indicated. }
\end{deluxetable}

\newpage
\tablenum{4}
\tablecolumns{8}
\tablewidth{0pt}

\begin{deluxetable}{lcccccccc}

\tablecaption{External Data Source Stellar Model Parameters and Individual [Fe/H] and [Mn/Fe] Values}
\tablehead{
\colhead{Association}                            &
\colhead{Star}                                   &
\colhead{T$_{eff}$}                              &
\colhead{log g}                                  &
\colhead{v$_t$}                                  &
\colhead{$[Fe/H]$}                               &
\colhead{$[Fe/H]$}                               &
\colhead{$[Mn/Fe]$}                             \\
\colhead{}                                       &
\colhead{}                                       &
\colhead{(K)}                                    &
\colhead{}                                       &
\colhead{(km s$^{-1}$)}                          &
\colhead{LIT}\tablenotemark{a}                   &
\colhead{}                                       &
\colhead{}                                       &
}
\startdata
Cr 261  &	1045	&	4400	&	1.50	&	1.20	&	$-$0.16	&	$-$0.14	&	$-$0.41	\\
Cr 261  &	1080	&	4490	&	2.20	&	1.20	&	$-$0.11	&	$-$0.25	&	$-$0.45	\\
Cr 261  &	1871	&	4000	&	0.70	&	1.50	&	$-$0.31	&	$-$0.59	&	$-$0.22	\\
Cr 261  &	2105	&	4300	&	1.50	&	1.50	&	$-$0.32	&	$-$0.47	&	$-$0.21	\\
NGC 288	&	20	&	4050	&	0.60	&	1.75	&	$-$1.44	&	$-$1.62	&	$-$0.31	\\
NGC 288	&	231	&	4300	&	1.10	&	1.50	&	$-$1.41	&	$-$1.50	&	$-$0.32	\\
NGC 288	&	245	&	4250	&	0.80	&	1.40	&	$-$1.41	&	$-$1.47	&	$-$0.30	\\
NGC 288	&	274	&	4025	&	0.70	&	1.90	&	$-$1.37	&	$-$1.48	&	$-$0.33	\\
NGC 288	&	281	&	4125	&	0.60	&	1.71	&	$-$1.42	&	$-$1.65	&	$-$0.29	\\
NGC 288	&	287	&	4350	&	1.20	&	1.40	&	$-$1.45	&	$-$1.34	&	$-$0.44	\\
NGC 288	&	297	&	4330	&	1.20	&	1.70	&	$-$1.41	&	$-$1.62	&	$-$0.23	\\
NGC 288	&	307	&	4350	&	1.20	&	1.35	&	$-$1.40	&	$-$1.63	&	$-$0.24	\\
NGC 288	&	338	&	4325	&	1.30	&	1.60	&	$-$1.37	&	$-$1.55	&	$-$0.28	\\
NGC 288	&	344	&	4180	&	0.80	&	1.60	&	$-$1.36	&	$-$1.45	&	$-$0.32	\\
NGC 288	&	351	&	4330	&	1.20	&	1.55	&	$-$1.30	&	$-$1.53	&	$-$0.36	\\
NGC 288	&	403	&	3950	&	0.20	&	1.90	&	$-$1.43	&	$-$1.59	&	$-$0.32	\\
NGC 288	&	531	&	3780	&	0.10	&	1.60	&	$-$1.31	&	$-$1.70	&	$-$0.42	\\
NGC 362	&	1137	&	4000	&	0.70	&	2.00	&	$-$1.37	&	$-$1.51	&	$-$0.31	\\
NGC 362	&	1159	&	4125	&	0.80	&	1.90	&	$-$1.27	&	$-$1.37	&	$-$0.34	\\
NGC 362	&	1334	&	3975	&	0.40	&	1.95	&	$-$1.30	&	$-$1.37	&	$-$0.42	\\
NGC 362	&	1401	&	3875	&	0.00	&	1.90	&	$-$1.32	&	$-$1.39	&	$-$0.34	\\
NGC 362	&	1423	&	3950	&	0.10	&	2.35	&	$-$1.37	&	$-$1.42	&	$-$0.41	\\
NGC 362	&	1441	&	3975	&	0.20	&	1.90	&	$-$1.31	&	$-$1.44	&	$-$0.29	\\
NGC 362	&	2115	&	3900	&	0.00	&	2.30	&	$-$1.38	&	$-$1.49	&	$-$0.32	\\
NGC 362	&	2127	&	4110	&	0.60	&	2.25	&	$-$1.30	&	$-$1.52	&	$-$0.39	\\
NGC 362	&	2423	&	4000	&	0.40	&	1.85	&	$-$1.32	&	$-$1.42	&	$-$0.41	\\
NGC 362	&	77	&	4075	&	0.20	&	2.50	&	$-$1.34	&	$-$1.41	&	$-$0.34	\\
NGC 362	&	MB2	&	4100	&	0.60	&	2.25	&	$-$1.30	&	$-$1.58	&	$-$0.20	\\
NGC 362	&	V2	&	3950	&	0.10	&	2.70	&	$-$1.30	&	$-$1.58	&	$-$0.48	\\
NGC 3201 &	5	&	4750	&	1.80	&	1.70	&	$-$1.38	&	$-$1.53	&	$-$0.42	\\
NGC 3201 &	8	&	4410	&	1.50	&	1.80	&	$-$1.17	&	$-$1.56	&	$-$0.28	\\
NGC 3201 &	9	&	4600	&	1.90	&	1.70	&	$-$1.18	&	$-$1.45	&	$-$0.34	\\
NGC 3201 &	42	&	4500	&	1.50	&	2.00	&	$-$1.32	&	$-$1.65	&	$-$0.29	\\
NGC 3201 &	112	&	4350	&	1.30	&	1.60	&	$-$1.38	&	$-$1.64	&	$-$0.18	\\
NGC 3201 &	121	&	4000	&	0.00	&	2.00	&	$-$1.40	&	$-$1.56	&	$-$0.42	\\
NGC 3201 &	168	&	4100	&	0.20	&	1.80	&	$-$1.42	&	$-$1.61	&	$-$0.30	\\
NGC 3201 &	238	&	4250	&	0.90	&	1.80	&	$-$1.42	&	$-$1.53	&	$-$0.44	\\
NGC 3201 &	293	&	4250	&	1.20	&	1.80	&	$-$1.39	&	$-$1.56	&	$-$0.38	\\
NGC 3201 &	301	&	4250	&	1.00	&	2.20	&	$-$1.49	&	$-$1.65	&	$-$0.33	\\
NGC 3201 &	312	&	4250	&	0.70	&	1.80	&	$-$1.47	&	$-$1.49	&	$-$0.49	\\
NGC 3201 &	318	&	4350	&	0.80	&	1.90	&	$-$1.52	&	$-$1.59	&	$-$0.39	\\
NGC 3201 &	357	&	4150	&	0.70	&	2.00	&	$-$1.55	&	$-$1.78	&	$-$0.21	\\
NGC 3201 &	419	&	4500	&	1.20	&	1.70	&	$-$1.28	&	$-$1.40	&	$-$0.58	\\
NGC 6287 &	1491 &	4375 &	1.00 &	1.75 &	$-$2.15 &	$-$2.28 &	$-$0.30 \\
NGC 6287 &	1387 &	4250 &	0.80 &	1.90 &	$-$2.10 &	$-$2.33 &	$-$0.26 \\
NGC 6293 &	2673 &	4250	&	0.50	&	1.90	&	$-$2.16	&	$-$2.24	&	$-$0.35	\\
NGC 6293 &	3857 &	4450	&	0.70	&	1.75	&	$-$2.18	&	$-$2.40	&	$-$0.39	\\
NGC 6528 &	I-42	&	4200	&	1.60	&	1.20	&	$-$0.14	&	$-$0.34	&	$-$0.23	\\
NGC 6528 &	I-36	&	4300	&	1.50	&	1.50	&	$-$0.13	&	$-$0.37	&	$-$0.21	\\
NGC 6528 &	I-18	&	4800	&	2.00	&	1.50	&	$-$0.05	&	$-$0.03	&	$-$0.33	\\
NGC 6541 &	I-44	&	4250	&	0.70	&	1.85	&	$-$1.85	&	$-$1.95	&	$-$0.28	\\
NGC 6541 &	II-113	&	4200	&	0.50	&	1.80	&	$-$1.86	&	$-$1.91	&	$-$0.36	\\
NGC 6705 (M11) &	660	&	4500	&	1.50	&	2.00	&	0.05 &	0.03	&	$-$0.43	\\
NGC 6705 (M11) &	669	&	4500	&	1.40	&	2.00	&	0.09 &	0.18	&	$-$0.32	\\
NGC 6705 (M11) &	686	&	4600	&	2.00	&	2.00	&	0.13 &	0.13	&	$-$0.45	\\
NGC 6705 (M11) &	779	&	4250	&	1.60	&	2.50	&    $-$0.01 &  $-$0.07	&	$-$0.49	\\
NGC 6705 (M11) &	916	&	4500	&	1.30	&	2.00	&	0.01 &	$-$0.04	&	$-$0.34	\\
NGC 6705 (M11) &	926	&	4500	&	1.50	&	2.00	&    $-$0.21 &	$-$0.59	&	$-$0.38	\\
NGC 6705 (M11) &	1184	&	4600	&	2.20	&	2.00	&	0.24 &	0.24	&	$-$0.30	\\
NGC 6705 (M11) &	1223	&	4750	&	2.50	&	2.00	&	0.19 &	$-$0.06	&	$-$0.39	\\
NGC 6705 (M11) &	1256	&	4600	&	2.50	&	2.00	&	0.31 &	0.28	&	$-$0.17	\\
NGC 6705 (M11) &	1423	&	4750	&	2.90	&	2.50	&	0.21 &	0.04	&	$-$0.05	\\
NGC 6752 &	1	&	4749	&	1.95	&	1.41	&	$-$1.58	&	$-$1.69	&	$-$0.43	\\
NGC 6752 &	2	&	4779	&	1.98	&	1.39	&	$-$1.59	&	$-$1.69	&	$-$0.48	\\
NGC 6752 &	3	&	4796	&	2.03	&	1.42	&	$-$1.64	&	$-$1.74	&	$-$0.39	\\
NGC 6752 &	4	&	4806	&	2.04	&	1.40	&	$-$1.61	&	$-$1.66	&	$-$0.46	\\
NGC 6752 &	6	&	4804	&	2.06	&	1.40	&	$-$1.61	&	$-$1.87	&	$-$0.38	\\
NGC 6752 &	7	&	4829	&	2.10	&	1.33	&	$-$1.84	&	$-$1.74	&	$-$0.53	\\
NGC 6752 &	8	&	4910	&	2.15	&	1.33	&	$-$1.62	&	$-$1.60	&	$-$0.43	\\
NGC 6752 &	9	&	4824	&	2.11	&	1.38	&	$-$1.63	&	$-$1.72	&	$-$0.41	\\
NGC 6752 &	10	&	4836	&	2.13	&	1.37	&	$-$1.60	&	$-$1.63	&	$-$0.52	\\
NGC 6752 &	11	&	4829	&	2.13	&	1.32	&	$-$1.64	&	$-$1.65	&	$-$0.47	\\
NGC 6752 &	12	&	4841	&	2.15	&	1.34	&	$-$1.62	&	$-$1.72	&	$-$0.40	\\
NGC 6752 &	15	&	4850	&	2.19	&	1.35	&	$-$1.61	&	$-$1.75	&	$-$0.45	\\
NGC 6752 &	16	&	4906	&	2.24	&	1.32	&	$-$1.60	&	$-$1.73	&	$-$0.44	\\
NGC 6752 &	19	&	4928	&	2.32	&	1.29	&	$-$1.61	&	$-$1.75	&	$-$0.49	\\
NGC 6752 &	20	&	4929	&	2.33	&	1.32	&	$-$1.59	&	$-$1.69	&	$-$0.48	\\
Pal 12 &	S1	&	3900	&	0.63	&	1.80	&	$-$0.76	&	$-$0.81	&	$-$0.31	\\
Pal 12 &	1118	&	4000	&	0.84	&	1.80	&	$-$0.80	&	$-$0.82	&	$-$0.35	\\
Pal 12 &	1128	&	4260	&	1.30	&	1.70	&	$-$0.82	&	$-$0.84	&	$-$0.40	\\
Pal 12 &	1305	&	4465	&	1.62	&	1.70	&	$-$0.80	&	$-$0.84	&	$-$0.38	\\
\enddata

\tablecomments{The complete version of this table is in the electronic
edition of the Journal.  The printed edition contains only a sample.}
\tablenotetext{a}{As discussed in text, literature values of [Fe/H] are provided.}
\end{deluxetable}

\newpage
\tablenum{5}
\tablecolumns{5}
\tablewidth{0pt}

\begin{deluxetable}{lccccccc}

\tablecaption{Line Parameters}
\tablehead{
\colhead{Element}                               &
\colhead{$\lambda$ [\AA]}                       &
\colhead{$\chi$ [eV]}                           &
\colhead{log (gf)}                              &
\colhead{E$\gamma$}                             \\
}
\startdata
Fe I & 6024.06 & 4.545 &  0.040    & 2.2\\
Fe I & 6027.05 & 4.073 & $-$1.089  & 2.0\\
Mn I & 6013.51 & 3.070 & $-$0.251  & 1.5\\
Mn I & 6016.64 & 3.071 & $-$0.216  & 1.5\\
Mn I & 6021.82 & 3.073 &  0.034    & 1.5\\
\enddata
\end{deluxetable}

\newpage
\tablenum{6}
\tablecolumns{10}
\tablewidth{0pt}
\tabletypesize{\scriptsize}
\begin{deluxetable}{lccccccccc}
\tablecaption{LTG Cluster Mean Abundances}
\tablehead{
\colhead{Cluster}                                &
\colhead{N$_{Stars}$}                            &
\colhead{$<$[Fe/H]$>$}                           &                                  
\colhead{$\sigma$}                               &
\colhead{$<$[Fe/H]$>$$_{LIT}$}                   &                                  
\colhead{$\sigma$}                               &
\colhead{[Fe/H]$_{AVG}$}\tablenotemark{a}        &
\colhead{$<$[Mn/Fe]$>$}                          &                                  
\colhead{$\sigma$}                               &
\colhead{$<$[Mn/Fe]$>$$_{AVG}$}\tablenotemark{b}\\
}
\startdata
NGC 5272 (M3)  & 20 & $-$1.59 & 0.08 & $-$1.59 & 0.07 & $-$1.59 & $-$0.43 & 0.09 & $-$0.43\\
NGC 5904 (M5)  & 31 & $-$1.37 & 0.11 & $-$1.34 & 0.08 & $-$1.35 & $-$0.40 & 0.08 & $-$0.41\\
NGC 6121 (M4)  & 20 & $-$1.24 & 0.06 & $-$1.19 & 0.02 & $-$1.21 & $-$0.39 & 0.07 & $-$0.42\\
NGC 6205 (M13) & 17 & $-$1.67 & 0.06 & $-$1.60 & 0.05 & $-$1.64 & $-$0.38 & 0.04 & $-$0.41\\
NGC 6254 (M10) & 12 & $-$1.64 & 0.11 & $-$1.53 & 0.09 & $-$1.58 & $-$0.38 & 0.08 & $-$0.43\\
NGC 6341 (M92) & 4  & $-$2.45 & 0.05 & $-$2.27 & 0.04 & $-$2.36 & $-$0.37 & 0.05 & $-$0.45\\
NGC 6838 (M71) & 10 & $-$1.13 & 0.15 & $-$0.79 & 0.05 & $-$0.96 & $-$0.16 & 0.14 & $-$0.32\\
NGC 7006       & 6  & $-$1.62 & 0.11 & $-$1.59 & 0.04 & $-$1.60 & $-$0.42 & 0.07 & $-$0.43\\
NGC 7078 (M15) & 11 & $-$2.51 & 0.05 & $-$-2.42 & 0.03 & $-$2.46 & $-$0.36 & 0.12 & $-$0.40\\  
Pal 5          & 4  & $-$1.51 & 0.10 & $-$-1.34 & 0.04 & $-$1.42 & $-$0.29 & 0.07 & $-$0.36\\
\enddata
\tablenotetext{a}{These values are the average of the [Fe I/H] values from this study and literature.}
\tablenotetext{b}{These values are computed using [Fe/H](AVG).}
\end{deluxetable}

\end{document}